\documentclass[10pt]{article} 

\usepackage[utf8]{inputenc}
\usepackage{amsmath}
\usepackage{amsfonts}
\usepackage{amssymb}
\usepackage{mathtools}
\usepackage{listings}
\usepackage{mathrsfs}
\usepackage{times}
\usepackage{float}
\usepackage{color}
\usepackage{setspace}
\usepackage{color,soul}

\usepackage{amsmath,amsfonts,amsthm,eucal}
\usepackage{enumerate}
\usepackage[letterpaper,margin=3cm]{geometry}
\usepackage{float}
\usepackage[title]{appendix}
\usepackage{color}
\usepackage{tabularx}
\usepackage{siunitx,stmaryrd}
\usepackage[tableposition=below]{caption}
\usepackage{lipsum}
\usepackage[FIGBOTCAP,TABTOPCAP,bf,tight]{subfigure}

\usepackage{natbib}
\usepackage[pdftex]{graphicx}
\usepackage{epstopdf}
\usepackage{booktabs} 
\usepackage{tikz,mathpazo}
\usetikzlibrary{shapes.geometric, arrows}
\usepackage{caption}

\newcommand{\tensor}[1]{\ensuremath{\boldsymbol{#1}}}

\DeclareMathOperator{\grad}{\nabla}

\theoremstyle{remark}

\renewcommand{\vec}[1]{\ensuremath{\boldsymbol{#1}}}

\usepackage{algorithm}
\usepackage{algorithmicx}
\usepackage[noend]{algpseudocode}

\theoremstyle{definition}

\usepackage{longtable}
\usepackage{adjustbox}
\usepackage{framed}

\usepackage{algorithm}
\usepackage[noend]{algpseudocode}

\title{ILS-MPM: an implicit level-set-based material point method for frictional particulate contact mechanics of deformable particles}

\begin{document}

\author{Chuanqi Liu\thanks{Department of Civil Engineering and Engineering Mechanics, 
 Columbia University, 
 New York, NY 10027.     \textit{cl3851@columbia.edu}  }       \and
        WaiChing Sun\thanks{Department of Civil Engineering and Engineering Mechanics, 
 Columbia University, 
 New York, NY 10027.
  \textit{wsun@columbia.edu}  (corresponding author)      
}
}

\maketitle

\begin{abstract}
Finite element simulations of frictional multi-body contact problems via conformal meshes can be challenging and computationally demanding. To render geometrical features, unstructured meshes must be used and this unavoidably increases the degrees of freedom and therefore makes the construction of slave/master pairs more demanding.
In this work, we introduce an implicit material point method designed to bypass the meshing of bodies by employing level set functions to represent boundaries at structured grids. This implicit function representation provides an elegant mean to link an unbiased intermediate reference surface with the true boundaries by closest point projection as shown in \citep{leichner2019contact}. We then enforce the contact constraints by a penalty method where the Coulomb friction law is implemented as an elastoplastic constitutive model such that a return mapping algorithm can be used to provide constitutive updates for both the stick and slip states.
To evolve the geometry of the contacts properly, the Hamilton-Jacobi equation is solved incrementally such that the level set and material points are both updated accord to the deformation field. To improve the accuracy and regularity of the numerical integration of the material point method, a moving least square method is used to project numerical values of the material points back to the standard locations for Gaussian-Legendre quadrature.
Several benchmarks are used to verify the proposed model. Comparisons with discrete element simulations are made to analyze the importance of stress fields on predicting the macroscopic responses of granular assemblies.
\end{abstract}

\section{Introduction}

Modeling frictional contacts in assemblies of deformable bodies has been an important subject of interest over the last several decades \citep{oden1983nonlocal, laursen1993formulation, de2011large}. Despite the compelling progress achieved in the subject, simulatio simulating the path-dependent responses of multiple evolving contacts remains a challenging task, especially for contacts involving multiple deformable bodies of non-convex and non-smooth geometries (e.g. sand, coal, slit, snow, powder). Nevertheless, understanding the grain-scale particulate frictional interactions is critical to analyze and predict macroscopic behaviors of particulate materials \citep{sun2013multiscale, hurley2014extracting, wang2019meta, wang2019updated}.

A popular way to handle the computational challenges of the multi-particle contact problems is to idealize the nature of the contact by allowing overlapping of particles and run the simulations with an explicit time integration scheme \citep{cundall1988formulation, wang2016semi, liu2016determining}. These treatments are implemented in the so-called discrete element method (DEM). The DEM bypasses the prediction of particle deformation and greatly reduces the total number of degrees of freedom (DOFs). However, the trade-off of this convenience is a less realistic prediction with the following shortcomings.

\begin{enumerate}
\item The local stress and deformation state of individual particle remain unknown since force is generated from overlapping particles or through rigid contacts \citep{rougier2004numerical}.

\item Any history-dependent behaviors of contacts can only manifest and be stored in particle-contact-pair, an edge of the connectivity graph, but not the particle surfaces. This simplification may be unrealistic for the surface with local friction, wear and lubrication.

\item The steady state of the particulate system must be obtained via relaxation or ad hoc damping \citep{liu2016nonlocal, rojek2007multiscale, modenese2012numerical}.

\item Non-spherical particles with complex geometries can be difficult to be represented \citep{houlsby2009potential, cho2006particle, andrade2012granular}.

\item The evolution of contact geometry due to particle deformation is either neglected or greatly simplified \citep{janko2018contact}.

\end{enumerate}

To capture the interactions of particles of non-spherical shapes, an implicit function representation approach (cf. \citet{houlsby2009potential, andrade2012granular}) is often used. First proposed in \citet{houlsby2009potential}, this numerical algorithm uses level set or signed distance function to identify the location of the particle boundaries. Contacts and the overlapping distance of the particles are then computed via a constrained optimization algorithm.
While these improved models are useful to locate the contact location, the only kinematic information stored remains the relative displacement of the overlapped particles in each contact pair. As such, the only kinetic data it can generate is the force and moment exerted at the contact pair.
This approach is therefore limited to the applications to the few cases where the geometrical nature of the contacts, the evolution of the stress and resultant nature of the history-dependence (e.g. frictional wear vs. bulk plasticity of particles) are negligible.

An alternative but computationally more demanding approach is to model the deformation of particles and capture the evolution of contact explicitly in a finite element (FE) framework. In fact, FE algorithms for frictional contact have been extensively studied over the past few decades. By designating master/slave (or mortar/non-mortar) pairs, contact constraints are imposed via node-to-segment \citep{wriggers2004computational, zavarise2009node} or segment-to-segment (latest mortar method) \citep{puso2004mortar, tur2009mortar, zimmerman2018surface}. These biased approaches are confronted with difficulties especially in the case of multi-body contact where it is impractical to a priori nominate a master surface and a slave one \citep{chouly2018unbiased}. Some unbiased formulations for contact were proposed to avoid these difficulties, for instance,
\citet{sauer2015unbiased, mlika2017unbiased}. Another important ingredient to complete the contact formulation is the method chosen to enforce contact constraints. While a penalty method may simply enforce the constraints by adding penalty or virtual power in the formulation \citep{belytschko2013nonlinear}, the tuning of the penalty parameter must be carried out to strike a dedicated balance between preventing overlapping and avoiding ill-conditioned tangential stiffness matrix for the implicit algorithm \citep{khoei2007enriched, liu2010stabilized}. The Lagrange multiplier method introduces extra unknown(s) to enforce the constraint exactly and therefore does not require tuning. However, the introduction of the additional governing
equation may lead to a saddle-point problem that is more difficult to solve \citep{tur2009mortar, fortin2013preconditioned}.
Based on the penalty and Lagrange multiplier methods, some variations of strategies can be developed. For instance, one may add extra terms to the total energy to improve the stabilization of calculation and conditioning of the matrix. This may lead to the augmented Lagrangian method (which introduces an additional functional to modify the effective stiffness) or the perturbed Lagrangian method (which introduces an additional functional to enable static condensation for the Lagrangian multiplier) \citep{belytschko2013nonlinear, tur2015modified}. The challenge with Lagrange multiplier methods for this class of problems concerns the construction of a stable Lagrange multiplier space to satisfy the inf-sup condition, which often requires the interpolation space for the Lagrange multiplier field to be coarsened with respect to the underlying mesh \citep{ji2004strategies, kim2007mortared,bechet2009stable, sun2014multiscale, sun2017mixed}.
Nitsche’s method was originally proposed in \cite{nitsche1971variationsprinzip} to enforce a Dirichlet boundary condition weakly. As a result of the pioneering work of \cite{hansbo2002unfitted}, Nitsche’s method has become popular for a wide class of contact problems.
This popularity could be attributed to the elegant treatment that eliminates both the outer augmentation loop as well as additional unknowns, i.e., the Lagrange multiplier, and therefore also eliminates the need for inf-sup stable mixed-field discretization \citep{annavarapu2014nitsche, mlika2017unbiased, chouly2018unbiased}.

Although the FM method can be applied to simulate contact problems, the generation and validating conformal meshes for finite element simulations remain a time-consuming process. For objects of complex geometries, conformal meshes may lead to local refinement that significantly increases the computational time but contributes little insight to the state of the assemblies.
To overcome this issue, the extended finite element method (X-FEM) \citep{moes1999finite, dolbow2001extended, liu2019modeling} or the generalized finite element method (G-FEM) \citep{melenk1996partition,simone2006generalized}
can be used to capture the interface geometry without a conformal unstructured mesh.  By selecting the enrichment function to span the solution space, this method can handle calculations with discontinuities provided that the specific nodal points are enriched via the addition of extra DOFs \citep{mousavi2010generalized}. The strategy of implementing phase field (PF) method to deal with frictional contact problems is almost identical to the enriched methods, except that the strong discontinuity is replaced by a regularized indicator function \citep{fei2019phase}.

Another strategy to bypass the need for conformal meshing is to use the immersed boundary (IB) elements where an underlying Cartesian structured grid consists of only regular hexahedral elements. The boundary is represented by partitioning the elements across the interfaces. Any integration involving the cut element is then only performed in the interior portion of the element \citep{zhang2004immersed}. Previous work, such as \citet{tur2015modified, ruberg2016unstructured}, has shown that it is possible to leverage the IB elements to reduce the total number of DOFs in the contact problems. \cite{leichner2019contact} develops a contact algorithm for voxel-based meshes enriched by the level set to handle the frictionless contacts.

The material point method (MPM), originally developed to deal with large deformation problems \citep{sulsky1994particle,sulsky1995application,zhou1999simulation, bardenhagen2004generalized}, is another approach that may solve contact problems without a conformal mesh. In the MPM, all physical information is carried by a set of material points. These points are connected with a background grid that regularly composed with structured cells. After solving the governing equations, the nodal solutions of the background mesh are updated and the movements of nodes are projected to the material points via the interpolation basis functions. The background mesh is then reset for the next incremental step. The implicit MPM assembles the global matrices in a similar fashion to the FE method counterpart where the material points work as integration points \citep{sulsky2004implicit}. However, since the locations of the material points can be arbitrary, the accuracy of the volume integration can be low and/or fluctuating in the temporal-spatial domain. \citet{sulsky2016improving} implemented an integration algorithm that employs fixed integration points in the MPM. The major idea is to project the data stored in the scattered material points back to the standard integration points via the moving least squares (MLS) method. The projected data is then used to compute integrals. This method is convenient for handling issues, such as hourglass mode or lockings. However, we need to pay attention to this competing mechanism since the accuracy would reduce due to the projection.

To handle contact problems, the MPM provides a convenient mean by aligning the normal direction of the contact surface with the gradient of mass or volume \citep{bardenhagen2000material, bardenhagen2001improved}.
However, the resolution of the interfaces depends on both the size of the background mesh and the number of material points. It is also difficult to cope with quasi-static contacts since contact is detected via existing relative velocities between two contacting bodies \citep{liu2018coupling}. As an alternative, we can implicitly represent the boundaries or interfaces via level set function defined on a Eulerian grid \citep{osher2004level}. \cite{zhang2017incompressible} enriched the level set function to the background mesh of the MPM to capture the evolution of the boundary. Moreover, the level set function is a versatile tool to help handle contacts since it provides an easy mean to compute the normal directions of the contact surfaces \citep{chi2015level}. The kernel particle method, as a meshfree method, also can handle contact problems within the framework of kernel functions \citep{sherburn2015meshfree, hillman2014stabilized, chen2017meshfree}.

In this work, our new contribution is two-fold. First, we introduce the level set representation of boundary for material point contact mechanics problems. As the implicit MPM allows one to keep track of the evolving interface without requiring small time step, this treatment provides a simple way to capture the interplays between the evolving geometry of the contact surfaces and the macroscopic frictional responses of the materials. Second, by introducing the level set representation of interfaces, the new implicit level set material point method, which we referred as ILS-MPM herein, is able to handle both the slip and stick states in the same manner while resolving the mesh-distortion and potential spurious oscillation through a projection technique inspired by the MLS approach used in \cite{bardenhagen2001improved}.

The rest of the paper is organized as follows. Section 2 states the problem setting giving the expression of total energy and the enforced constraints for frictional contact problems. Section 3 shows the formulations related to the level set, such as the evolution equation, the extension of physical fields and the gap function in the context of the level set. Section 4 presents the spatial discretization scheme and the temporal integration algorithm, focusing on the algorithm of the MPM. Section 5 gives some verifications and numerical examples to show the effectiveness of the proposed method to simulate granular materials. Section 6 concludes the paper by the major remarks.

\section{Problem setting}
In this section, we start with the expression of total energy of contact problems and the required constraints. The energy contributions arising from contact tractions are then derived in analogue to the Nitsche's method to circumvent the extra DOFs introduced in the Lagrangian multiplier formulation.

\subsection{Total energy in a Lagrangian multiplier formulation}
\begin{figure}[H]
\centering
\includegraphics[width=8cm]{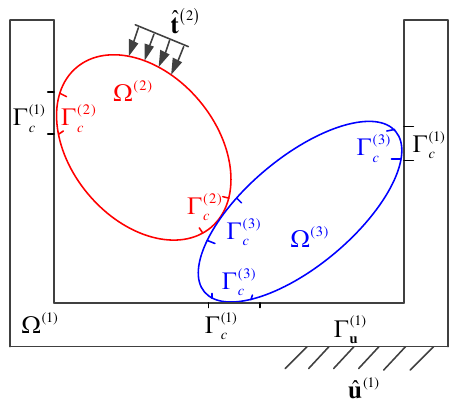}
\caption{Schematic of contact problems}
\label{figure_schematic}
\end{figure}

We consider contact problems involving finite solid bodies $\Omega^{(i)}$ with boundaries of $\partial \Omega^{(i)} $, where $i \in \mathcal{I} :=\{1,\cdots, N_\Omega\}$ index the solids. There exists a finite bounding box $\mathcal{B}$ containing all $\Omega^{(i)}$, and we may write:
\begin{equation}
\bigcup_{i\in\mathcal{I}} \Omega^{(i)} =: \Omega \subset \mathcal{B} \subset \mathbb{R}^d,
\end{equation}
where $d$ is the number of spatial dimensions. Similar to $\Omega$, we can define $\partial \Omega \subset \mathbb{R}^{d-1} $ as a union of $\partial \Omega^{(i)}$, and assume $\partial \Omega = \Gamma_{d} \cup \Gamma_{n} \cup \Gamma_{c}$ with $\Gamma_{d} \cap \Gamma_{n} = \emptyset$, $\Gamma_{d} \cap \Gamma_{c} = \emptyset$ and $\Gamma_{n} \cap \Gamma_{c} = \emptyset$, where $\Gamma_{d}$, $\Gamma_{n}$, and $\Gamma_{c}$ represent the Dirichlet, Neumann and contact boundary surfaces, respectively. We denote $\vec{u} = \hat{\vec{u}}$ on $\Gamma_{d}$ and $\tensor{\sigma} \cdot \vec{n} = \hat{\vec{t}}$ on $\Gamma_{n}$, where $\vec{u}$ is the displacement, $\tensor{\sigma}$ is the stress tensor, $\vec{n}$ is the normal direction of the surface, $\hat{\vec{u}}$ is the prescribed displacement, and $\hat{\vec{t}}$ is the prescribed traction. Fig. \ref{figure_schematic} shows a schematic of contact problems.

The weak form of contact problems can be derived from the minimization of the total energy $\mathcal{W}$ with certain constraints \citep{tur2009mortar}. In the Lagrangian multiplier formulation, the total energy for the contact problems is defined as the sum of potential energy $\pi_p(\vec{u})$ and the contact contributions:
\begin{equation}
\label{eq:lagrangian}
\mathcal{W}(\vec{u}, \lambda_n, \lambda_t) := \pi_p(\vec{u}) + \pi_n(\vec{u}, \lambda_n) + \pi_{t}(\vec{u}, \lambda_{t}),
\end{equation}
where $\lambda_n$ and $\lambda_{t}$ are the Lagrangian multipliers along normal and tangential directions of the contact surface, respectively, $\pi_n$ and $\pi_{t}$ are contact contributions arising from normal and tangential tractions, respectively. Prior to considering the specific expressions of energy contributions, we discuss the constraints should be fulfilled for the contacts, which are called Karush-Kuhn-Tucker (KKT) conditions \citep{laursen2013computational}:

(1) normal direction
\begin{equation}
\label{eq:condition1}
g_n(\vec{u}) \ge 0, \quad \tau_n (\vec{u}) \le 0 ,\quad \tau_n (\vec{u}) g_n(\vec{u}) = 0 \quad \mathrm{on}~\Gamma_c,
\end{equation}

(2) tangential direction
\begin{equation}
\label{eq:condition2}
|\tau_t (\vec{u})| \le \mu |\tau_n (\vec{u})| \quad \mathrm{on}~\Gamma_{c},
\end{equation}
and
\begin{equation}
\label{eq:condition3}
g_t(\vec{u}) = \gamma \tau_t(\vec{u}),~{\rm{where}}~
\left\{
\begin{aligned}
\gamma = 0,~{\rm{if}}~ |\tau_t (\vec{u})| < \mu |\tau_n (\vec{u})| \\
\gamma \geq 0,{~\rm{if}}~|\tau_t (\vec{u})| =\mu |\tau_n (\vec{u})|
\end{aligned}
\right.,
\end{equation}
where $g_n$ is the normal gap function, $\tau_n = \vec{\tau} \cdot \vec{n}$ and $\tau_t = \vec{\tau} \cdot \vec{t}$ are normal and tangential components of the contact traction, respectively, $\vec{\tau}$ is the traction vector applied on the contact surface, $\vec{n}$ and $\vec{t}$ are normal and tangential directions, respectively, $g_t$ is the tangential slip, $\mu$ is the Coulomb coefficient of friction and $\gamma$ is a nonnegative coefficient. We briefly interpret the KKT conditions as follows.
\begin{enumerate}
\item For condition (\ref{eq:condition1}), the first term describes the non-penetration condition for any contact, the second term ensures that only inward contact forces act over the contact area, and the third term ensures that if there is contact (contact pressure is non-zero) the global gap is zero.
\item Condition (\ref{eq:condition2}) requires that the magnitude of the tangential stress not exceed the coefficient of friction times the contact pressure. Condition (\ref{eq:condition3}) represents two important physical ideas associated with the Coulomb law, first, that the tangential slip $g_t(\vec{u})$ be identically zero when the tangential stress is less than the Coulomb limit, and second, that any tangential slip that does occur be colinear with the frictional stress exerted by the sliding point on the opposing surface.
\end{enumerate}

\subsection{Contact contributions obtained by the penalty method}
As shown in (\ref{eq:lagrangian}), two extra DOFs, i.e., $\lambda_n$ and $\lambda_t$, are introduced in the Lagrangian formulation. We thus need to choose suitable functional spaces to ensure the inf-sup condition for the multiple fields \citep{bechet2009stable}, which complicates the computation. Therefore, we here analogize the Nitsche's method to compute the contact contributions without introducing extra DOFs, which is identical to the penalty method. For clarity, we only consider the contact between $\Omega^{(i)}$ and $\Omega^{(j)}$ and show the subscripts when it is necessary.

The normal contribution $\pi_n$ is first defined depending on a perturbed Lagrangian method as :
\begin{equation}
\label{eq:pi_n0}
\pi_n(\vec{u}, \lambda_n) = \int_{\Gamma_c} g_n(\vec{u}) \lambda_n d\gamma - \frac{\epsilon_n}{2} \int_{\Gamma_c} \lambda_n^2  d\gamma,
\end{equation}
where $\epsilon_n$ is a coefficient. For the saddle point problem, it then follows:
\begin{equation}
\frac{\partial \pi_n(\vec{u}, \lambda_n)} {\partial \lambda_n} = \int_{\Gamma_c} \left[ g_n(\vec{u}) - \epsilon_n \lambda_n \right] d\gamma = 0.
\end{equation}
We thus can assume:
\begin{equation}
\label{eq:tau_n}
\tau_n(\vec{u}) = \lambda_n(\vec{u}) = \frac{1} {\epsilon_n}g_n(\vec{u}).
\end{equation}
Substituting (\ref{eq:tau_n}) into (\ref{eq:pi_n0}), we final obtain the normal contribution without introducing extra DOFs as:
\begin{equation}
\label{eq:pi_n}
\pi_n(\vec{u}) = \frac{1}{2\epsilon_n} \int_{\Gamma_c} (g_n(\vec{u}))^2d\gamma,
\end{equation}
which is identical to the penalty method.

\begin{figure}[h]
\centering
\includegraphics[width=8cm]{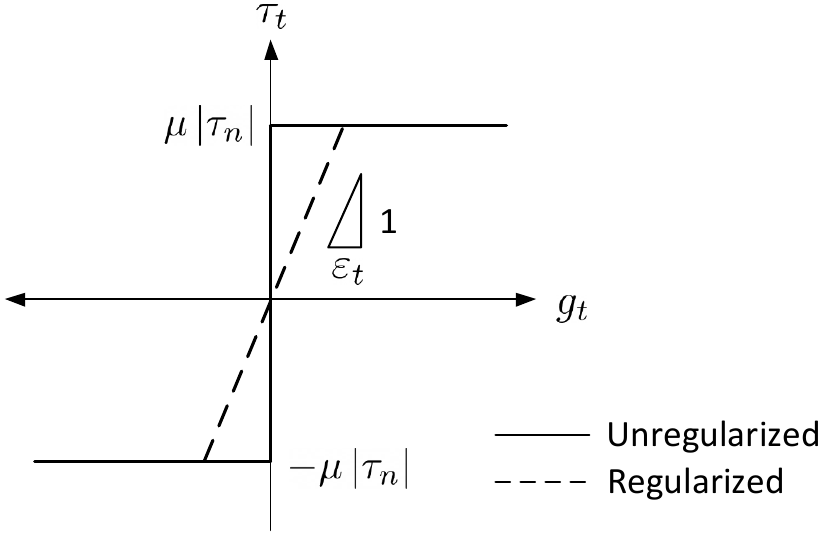}
\caption{Schematic depiction of the penalty regularized Coulomb friction law. }
\label{figure_normalized}
\end{figure}

In the tangential direction, (\ref{eq:condition2}) and (\ref{eq:condition3}) represent unregularized constraints implying that the relationship between $g_t$ and $\tau_t$ is a strong discontinuous function, which can cause zig-zagging in practice. As shown in Fig. \ref{figure_normalized}, following \cite{laursen2013computational}, we additively decompose the tangential slip into an "elastic" or recoverable part and a non-recoverable plastic part, such that:
\begin{equation}
\label{eq:gt}
g_t(\vec{u}) = g^e_t(\vec{u}) + g^p_t(\vec{u}).
\end{equation}
Therefore, we regularize the tangential constraints to a weak discontinuous function avoiding zig-zagging. Since the Coulomb friction law behaves analogously to an elastoplastic material model, we can consider the physical ideas reflected by (\ref{eq:condition2}) and (\ref{eq:condition3}), writing a related constitutive law as:
\begin{equation}
\label{eq:yield}
\begin{aligned}[r]
\phi(\tau_t, \tau_n) := |\tau_t (\vec{u})| - \mu | \tau_n(\vec{u}) | \leq 0, \\
\dot{g}_t = \dot{\gamma} \frac{\tau_t (\vec{u})}{|\tau_t (\vec{u})|}, \quad
\dot{\gamma} \geq 0, \quad \dot{\gamma} \phi = 0, \\
\dot{\gamma} \dot{\phi} = 0~(\mathrm{if}~ \phi = 0),
\end{aligned}
\end{equation}
where $\phi$ is termed the slip function and is a direct analogue of the yield function in theories of plasticity. The second line in (\ref{eq:yield}) express the colinearity of slip displacement $g_t$ and frictional stress $\tau_t$ in rate form, and $\dot{\gamma}$ represents the slip rate. The third line in (\ref{eq:yield}) is the persistency condition to ensure that if elastic unloading begins to occur while a point is still on the yield surface, the plastic strain rate will be zero. The numerical benefit of the analogy to elastoplasticity is that it enables an appeal to the return mapping strategies leading an unified form to compute the energy contribution of tangential traction for all conditions of stick and slip.

In analogy to the derivations of (\ref{eq:tau_n}) and (\ref{eq:pi_n}), we thus can obtain the tangential traction and the corresponding contribution to the energy as:
\begin{equation}
\label{eq:tau_t}
\tau_{t}(\vec{u}) = \lambda_{t}(\vec{u}) = \frac{1} {\epsilon_t} (g_t(\vec{u}) - g^p_t(\vec{u})),
\end{equation}
and
\begin{equation}
\label{eq:pi_st}
\pi_{t}(\vec{u}) = \frac{1}{2\epsilon_t} \int_{\Gamma_{c}} (g_t(\vec{u}) - g^p_t(\vec{u}))^2d\gamma,
\end{equation}
where $\epsilon_t$ is a coefficient representing the elastic stiffness, as shown in Fig. \ref{figure_normalized}. Again, (\ref{eq:tau_t}) and (\ref{eq:pi_st}) are identical to the expressions shown in the penalty method. The above treatment of the Coulomb friction law is called the penalty regularization \citep{laursen2013computational}.

It is worthy noted that the so-called Nitsche's stabilized method \citep{bechet2009stable} derived from a modified perturbed Lagrangian method \citep{tur2015modified,leichner2019contact} as :
\begin{equation}
\label{eq:pi_n00}
\pi_n(\vec{u}, \lambda_n) = \int_{\Gamma_c} g_n(\vec{u}) \lambda_n d\gamma - \frac{\epsilon_n}{2} \int_{\Gamma_c} (\lambda_n - \sigma_n(\vec{u}))^2  d\gamma,
\end{equation}
where $\sigma_n(\vec{u}) = \vec{n} \cdot (\beta^{(i)} \tensor{\sigma}^{(i)} + \beta^{(j)} \tensor{\sigma}^{(j)} ) \cdot \vec{n}$, with $\beta^{(i)} + \beta^{(j)} = 1$, is the extension of normal stress at $\Gamma_c$ computed from the stress field of the bodies \citep{wriggers2004computational}. It can be derived that:
\begin{equation}
\label{eq:tau_n0}
\tau_n(\vec{u}) = \lambda_n(\vec{u}) = \sigma_n(\vec{u}) + \frac{1} {\epsilon_n}g_n(\vec{u}).
\end{equation}
Compared to (\ref{eq:tau_n}), the normal traction shown as (\ref{eq:tau_n0}) is partly contributed from the body stress. Hence, a moderate coefficient in (\ref{eq:tau_n0}) suffices the non-penetration constraint, rather than a huge penalty coefficient in (\ref{eq:tau_n}). However, for the frictional contact, the tangential traction in the Nitsche's stabilized method is:
\begin{equation}
\label{eq:tau_t0}
\tau_{t}(\vec{u}) = \lambda_{t}(\vec{u}) = \sigma_t(\vec{u}) + \frac{1} {\epsilon_t} (g_t(\vec{u}) - g^p_t(\vec{u})),
\end{equation}
which is not a typical elastic relation as shown in (\ref{eq:tau_t}). In practice, we confront convergence issues in a Newton-Raphson iteration scheme using the return-mapping algorithm to update the tangential traction defined in (\ref{eq:tau_t0}). That is why we here adopt the  penalty method.

In a summary, substituting (\ref{eq:pi_n}) and (\ref{eq:pi_st})  into (\ref{eq:lagrangian}), we can derive the total energy without Lagrangian multipliers. The main problems need to be solved now are how to define the gap functions and what is the contact surface. We introduce the level set method to answer these questions in the following section.

\section{Level set method for boundary tracking}
Since we use non-conformal meshes for the spatial discretization, the true boundaries of bodies are implicitly represented by level sets. In this section, we first present the definition of level set, and the corresponding governing equation. We then demonstrate how we can adopt the level set approach introduced by \cite{leichner2019contact} for frictionless contacts to frictional contacts in the proposed material point method. The core is to construct a potential contact reference without a priori nomination of a master surface and the slave counterpart. We then define the gap functions in the context of the level set.

\subsection{Representation of boundaries via level set}
A level set $\Phi(\vec{x})$ is expressed in the Eulerian coordinates to implicitly represents boundary surfaces. In a bounded domain $\mathcal{B}$, the level set reads,
\begin{equation}
\Phi: \mathcal{B} \rightarrow \mathbb{R}, ~
\Phi (\vec{x}) \left\{
\begin{aligned}
&< 0 \quad \vec{x} \in \Omega \\
&= 0 \quad \vec{x} \in \partial \Omega\\
&> 0 \quad \vec{x} \in \mathcal{B} / \bar{\Omega}\\
\end{aligned}
\right.,
\end{equation}
where $\bar{\Omega} = \Omega \cup \partial \Omega$ constitutes the interior body and boundary, and $\mathcal{B} / \bar{\Omega}$ represents the exterior domain. The outward-pointing unit normal on the boundary surface is easily obtained by:
\begin{equation}
\label{eq:n}
\vec{n}(\vec{x}) = \frac{\grad{\Phi(\vec{x})}}{||\grad{\Phi(\vec{x})}||}, \; \vec{x} \in \partial \Omega,
\end{equation}
where $||\cdot||$ is the norm operator and $\grad{}$ is the gradient operator. The evolution of level set is governed by the convection equation:
\begin{equation}
\label{eq:phit}
\Phi_t + \vec{v} \cdot \grad{\Phi} = 0,
\end{equation}
where the $t$ subscript denotes a temporal partial derivative in the time variable $t$ and $\vec{v}$ is the velocity vector. \cite{osher2004level} shows that (\ref{eq:phit}) is valid for $\forall \vec{x} \in \mathcal{B}$ when the $\vec{v}$ is smooth without sharp gradients, which is assumed in this work. For quasi-static problems, we can set $\vec{v} = \vec{u}$ by assuming the pseudo time interval between two loading steps is unit. Since we only can compute $\vec{u} (\vec{x}) $ for $\vec{x} \in \Omega$, an extension algorithm of $\vec{u}(\vec{x})$ to $\mathcal{B}/ \bar{\Omega}$ is needed to evolve the level set. The widely employed algorithm for the field extrapolation of a scalar field $f(\vec{x})$ is to carry out a constant normal extrapolation by solving the partial differential equation (PDE) to steady state:
\begin{equation}
\label{eq:extension}
f_t + \vec{n} \cdot \grad{f} = 0.
\end{equation}
In practice, this procedure can be carried out using a fast marching method (FMM), in which the field values are set by using $\vec{n}\cdot\grad{f}=0$. Details of the algorithm of the FMM can be found in \cite{osher2004level, Chopp2012}. Once we achieve $\vec{u}({\vec{x}})$ for $\forall \vec{x} \in \mathcal{B}$, we need a stable algorithm to solve (\ref{eq:phit}) and extensive efforts have been made for such an algorithm in the past decades. We here implement the Hamilton-Jacobi weight essentially non-oscillatory (HJ WENO) scheme \citep{jiang2000weighted} to discretize the spatial terms to fifth-order accuracy and total variation diminishing (TVD) Runge-Kutta (RK) methods \citep{shu1988efficient} to increase the accuracy of temporal discretization. A review of level set methods and some recent applications can be found in \cite{gibou2018review}.

Note that \cite{aslam2004partial} proposed an algorithm for higher-order field extrapolation and \cite{rycroft2012simulations} came up an algorithm to update the level set without solving \eqref{eq:phit} based on the high order field extrapolation and the FMM. However, for convenience, we here adopt an open-source code "Level Set Method Library (LSMLIB)" \citep{chu2008level} to track the evolution of boundaries.

\subsection{Construction of the candidate contact surfaces}
An unbiased candidate contact surface can be conveniently inferred from the level set as illustrated in \citet{chi2015level, leichner2019contact}. For completeness, we provide a brief account on the procedure that constructs the contact reference. More details on similar level set procedure can be found in their work. As an instance, we still consider the contact between $\Omega^{(i)}$ and $\Omega^{(j)}$.

\begin{figure}[H]
\centering
\includegraphics[width=12cm]{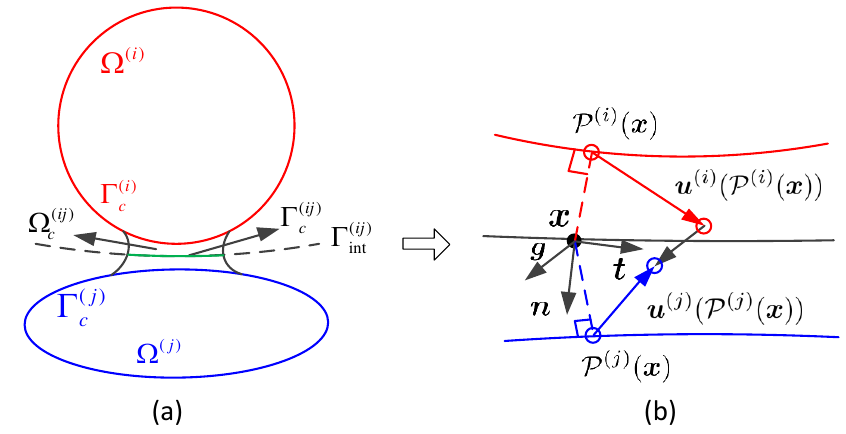}
\caption{Contact detecting: (a) tight proximity zone and (b) standard gap vector.}
\label{figure_contact}
\end{figure}

We first define the minimum level set field as:
\begin{equation}
\Phi^{(ij)}_{\rm{min}}(\vec{x}) = \left\{
\begin{aligned}
&\Phi^{(i)}(\vec{x}) \quad {\rm{if}}~\Phi^{(i)}(\vec{x}) < \Phi^{(j)}(\vec{x}) \\
&\Phi^{(j)}(\vec{x}) \quad \rm{otherwise}
\end{aligned}
\right..
\end{equation}
The region indicting the contact is specified:
\begin{equation}
\Omega^{(ij)}_{c} := \left\{ \left. \vec{x} ~\right| \Phi^{(i)}(\vec{x}) > 0 \cap ~ \Phi^{(j)}(\vec{x}) > 0 \cap ~ {\rm{reinitialize}}(\Phi^{(ij)}_{{\rm{min}}}(\vec{x}) - \epsilon) + \epsilon < 0\right\},
\end{equation}
where "reinitialize" refers to the procedure that converts the level set into a signed distance function
 (cf. \citet{osher2003signed, li2005level, sun2011multiscale}). $\epsilon$ is a shift parameter. As shown in Fig. \ref{figure_contact}a, $\Omega^{(ij)}_{c}$ often resembles a  liquid bridge between two potentially contacting particles. We further construct an intermediate level set $\Phi^{(ij)}_{\rm{int}}(\vec{x})$ whose zero-isocontour surface $\Gamma^{(ij)}_{\rm{int}}$ is used  as a reference for contact. The intermediate level set is defined by
\begin{equation}
\Phi^{(ij)}_{\rm{int}}(\vec{x}) = \frac{\Phi^{(i)}(\vec{x}) - \Phi^{(j)}(\vec{x})}{2}.
\label{eq:phiij}
\end{equation}
Eq. \eqref{eq:phiij} indicates that $\Phi^{(i)}(\vec{x}) = \Phi^{(j)}(\vec{x})$ for
 $\vec{x} \in \Gamma^{(ij)}_{\rm{int}}$ (i.e., $\Phi^{(ij)}_{\rm{int}}(\vec{x}) = 0$). Then, the following portion of the intermediate surface offers an alternative to the master and slave surfaces:
\begin{equation}
\label{eq:gammac}
\Gamma^{(ij)}_c := \Gamma^{(ij)}_{\rm{int}} \cup \Omega^{(ij)}_{c}.
\end{equation}
This surface (the green line in Fig. \ref{figure_contact}a) is an initial guess for a contact surface between $\Omega^{(i)}$ and $\Omega^{(j)}$. In the previous configuration,  $\Gamma^{(ij)}_c $ acts as a reference surface which supports the active set strategy \citep{laursen2013computational}.
Note that $\Gamma^{(ij)}_c$ is a fictitious interface aligned with neither surfaces.
However, in the updated deformed configuration, $\Gamma^{(ij)}_c$, $\Gamma^{(i)}_c$ and $\Gamma^{(j)}_c$ coincide when a new equilibrium state is achieved. At the conclusion of this step, we setup a contact reference for $\Omega^{(i)}$ and $\Omega^{(j)}$, $\Gamma_{c}^{(ij)}$ and generate a set of integration points along $\Gamma_{c}^{(ij)}$ for the active set strategy. 

\subsection{Determination of the gap function}
In order to enforce the contact constraints, such as non-penetration, we require a gap function that indicates the distance of boundaries of potentially contacting bodies. For the contact reference surface $\Gamma_{c}^{(ij)}$, any point on this surface $\vec{x} \in \Gamma_{c}^{(ij)}$ can be mapped onto the true boundary of the body $\partial \Omega^{(i)}$ via a closest point projection $\mathcal{P}^{(i)}(\vec{x}):\Gamma_{c}^{(ij)} \rightarrow \partial \Omega^{(i)}$, such that
$\Phi^{(i)} (\mathcal{P}^{(i)}(\vec{x})) = 0$ and $(\mathcal{P}^{(i)}(\vec{x}) - \vec{x}) \cdot \grad \Phi^{(i)} (\mathcal{P}^{(i)}(\vec{x})) = 0$. This projection can be achieved by the following iterative algorithm by setting $\vec{x}^0 \in \Gamma_{c}^{(ij)}$ \citep{rycroft2012simulations}:
\begin{equation}
\begin{aligned}
&\vec{\delta}_1 = -\Phi^{(i)}(\vec{x}^k)\frac{\grad \Phi^{(i)}(\vec{x}^k)}{\grad \Phi^{(i)}(\vec{x}^k) \cdot \grad \Phi^{(i)}(\vec{x}^k)}, \\
&\vec{x}^{k+1/2} = \vec{x}^k + \vec{\delta}_1, \\
&\vec{\delta}_2 = (\vec{x}^0 - \vec{x}^k) -\frac{(\vec{x}^0 - \vec{x}^k) \cdot\grad \Phi^{(i)}(\vec{x}^k) }{\grad \Phi^{(i)}(\vec{x}^k) \cdot \grad \Phi^{(i)}(\vec{x}^k)}\grad \Phi^{(i)}(\vec{x}^k), \\
& \vec{x}^{k+1} = \vec{x}^{k+1/2} + \vec{\delta}_2,
\end{aligned}
\end{equation}
where $\mathcal{P}^{(i)}(\vec{x}) = \vec{x}^n, n\to\infty$. We thus can define a gap vector for two contacting bodies with regard to $\vec{x} \in \Gamma^{(ij)}_c$ in the current configuration, as:
\begin{equation}
\label{eq:gapvector}
\vec{g}(\vec{u}^{(i)} (\mathcal{P}^{(i)}(\vec{x})), \vec{u}^{(j)} (\mathcal{P}^{(j)}(\vec{x})) ) := \mathcal{P}^{(j)}(\vec{x}) - \mathcal{P}^{(i)}(\vec{x}) + \vec{u}^{(j)} (\mathcal{P}^{(j)}(\vec{x})) - \vec{u}^{(i)} (\mathcal{P}^{(i)}(\vec{x})), 
\end{equation}
where $\vec{u}^{(i)}: \Omega^{(i)} \rightarrow \mathbb{R}^{d}$ is the displacement field of the $i$-th body. 
As shown in Fig. \ref{figure_contact}b, the normal direction of the contact at $\vec{x} \in \Gamma^{(ij)}_c$ is defined as the gradient of $\Phi^{(ij)}_{\rm{int}}$ via (\ref{eq:n}) and denoted to  $\vec{n}$. The tangential direction $\vec{t}$ can therefore be easily determined. In order to consider the linearity of functional, the normal gap is split into
\begin{equation}
\label{eq:gn}
g_n (\vec{u}^{(i)} (\mathcal{P}^{(i)}(\vec{x})), \vec{u}^{(j)} (\mathcal{P}^{(j)}(\vec{x}))) = g_{n0}(\vec{x}) + j^{(ij)}_{\rm{int}} (\vec{u}(\vec{x})), \\
\end{equation}
where $g_{n0}(\vec{x})$ represents the normal gap in the undeformed configuration and $j^{(ij)}_{\rm{int}} (\vec{u}(\vec{x}))$ is the normal jump, which are defined as:
\begin{align}
\label{eq:gn0}
g_{n0}(\vec{x}) &= (\mathcal{P}^{(j)}(\vec{x}) - \mathcal{P}^{(i)}(\vec{x})) \cdot \vec{n}, \\
\label{eq:j}
j^{(ij)}_{\rm{int}} (\vec{u}(\vec{x})) &= (\vec{u}^{(j)} (\mathcal{P}^{(j)}(\vec{x})) - \vec{u}^{(i)} (\mathcal{P}^{(i)}(\vec{x}))) \cdot \vec{n}.
\end{align}
In contrast to $g_n$, the tangential slip $g_t$ is the relative displacement in tangential direction as:
\begin{equation}
\label{eq:s}
g_t(\vec{u}^{(i)} (\mathcal{P}^{(i)}(\vec{x})), \vec{u}^{(j)} (\mathcal{P}^{(j)}(\vec{x}))) = (\vec{u}^{(j)} (\mathcal{P}^{(j)}(\vec{x})) - \vec{u}^{(i)} (\mathcal{P}^{(i)}(\vec{x}))) \cdot \vec{t}.
\end{equation}
For clarity, we also define a shear jump $s^{(ij)}_{\rm{int}}(\vec{u}(\vec{x}))$ in analogy to $j^{(ij)}_{\rm{int}}(\vec{u}(\vec{x}))$ as:
\begin{equation}
\label{eq:sj}
s^{(ij)}_{\rm{int}}(\vec{u}(\vec{x})) : = g_t(\vec{u}^{(i)} (\mathcal{P}^{(i)}(\vec{x})), \vec{u}^{(j)} (\mathcal{P}^{(j)}(\vec{x}))) - g^p_t(\vec{u}^{(i)} (\mathcal{P}^{(i)}(\vec{x})), \vec{u}^{(j)} (\mathcal{P}^{(j)}(\vec{x}))).
\end{equation}
where $g^p_t$ is the non-recoverable plastic part as shown in (\ref{eq:gt}).

\subsection{Active set strategy}
Since we analogize the Coulomb friction law to an elastoplastic constitutive law, it is not required to distinguish the tangential state, i.e., stick or slip explicitly. We only need to activate the points with negative normal traction $\tau_n$ to update the constitutive responses properly. 
The set of active points for contact between $\Omega^{(i)}$ and $\Omega^{(j)}$ is denoted as:
\begin{equation}
\mathcal{A}^{(ij)} := \{\vec{x} \in \Gamma^{(ij)}_c | \tau^{(ij)}_n(\vec{x}) < 0\}.
\end{equation}
where $\Gamma^{(ij)}_c$ is defined in (\ref{eq:gammac}) and $\tau^{(ij)}_n(\vec{x})$ is defined at (\ref{eq:tau_n}).

\section{Variations, discretization and integration}
We have derived the expression of total energy as shown above. In this section, we first show the procedure to derive the weak form of frictional contact problems via the first variation of the energy. We then present the spatial discretization for one body and the update algorithm of the MPM using the MLS method. The Newton-Raphson iteration scheme and the Jacobian matrix computed via finite-differencing of the residuals to converge the computations are given at end.

\subsection{Variations and weak form}
The first variation of $\pi_p(\vec{u})$ is
\begin{equation}
\label{eq:varpi_p}
\delta \pi_p(\vec{u}, \vec{v}) = a(\vec{u},\vec{v}) +  f(\vec{v})
\end{equation}
where $\vec{v}$ is the variation of displacement, and the bilinear forms $a(\cdot, \cdot)$ and linear operator $f(\cdot)$ are defined as:
\begin{equation}
a(\vec{u},\vec{v}) = \int_\Omega \tensor{\sigma}(\vec{u}) : \tensor{\varepsilon}(\vec{v}) d\vec{x} \quad f(\vec{v}) := \int_\Omega \vec{b} \cdot \vec{v} d\vec{x} + \int_{\Gamma_n} \hat{\vec{t}} \cdot \vec{v} d\gamma,
\end{equation}
where $\tensor{\varepsilon}$ is strain tensor and $\vec{b}$ is body force.

Considering (\ref{eq:gn}) and (\ref{eq:sj}), the first variations of (\ref{eq:pi_n}) and (\ref{eq:pi_st}) are:
\begin{equation}
\label{eq:varpi_n}
\delta \pi_n(\vec{u}, \vec{v}) = \frac{{\rm{d}}}{{\rm{d}}\varepsilon} \pi_n(\vec{u} + \varepsilon \vec{v})|_{\varepsilon = 0} = \int_{\Gamma_c} j^{(ij)}_{\rm{int}} (\vec{v}) \tau_n(\vec{u}) d\gamma,
\end{equation}
and
\begin{equation}
\label{eq:varpi_st}
\delta \pi_{t}(\vec{u}, \vec{v}) = \int_{\Gamma_{c}} s^{(ij)}_{\rm{int}} (\vec{v}) \tau_{t}(\vec{u}) d\gamma,
\end{equation}
respectively. Combining (\ref{eq:varpi_p}), (\ref{eq:varpi_n}) and  (\ref{eq:varpi_st}), the weak form for the contact problems is: find $\vec{u} \in H^1(\Omega)$, such that
\begin{equation}
\label{eq:weak}
r(\vec{u},\vec{v}) = a(\vec{u},\vec{v}) +
\int_{\Gamma_c} j^{(ij)}_{\rm{int}} (\vec{v}) \tau_n(\vec{u}) d\gamma  +
\int_{\Gamma_{c}} s^{(ij)}_{\rm{int}} (\vec{v}) \tau_{t}(\vec{u}) d\gamma
+ f(\vec{v}) = 0, ~ \forall \vec{v} \in H^1_0(\Omega).
\end{equation}
For each time step, we solve (\ref{eq:weak}) by using the Newton-Raphson iteration scheme. Before discussing the convergence strategy, we would like give some details on spatial discretization and updating procedure for each particular body under the framework the MPM considering the MLS method.

\subsection{The material point method implementing the MLS method}
\begin{figure}[h]
\centering
\includegraphics[width=16cm]{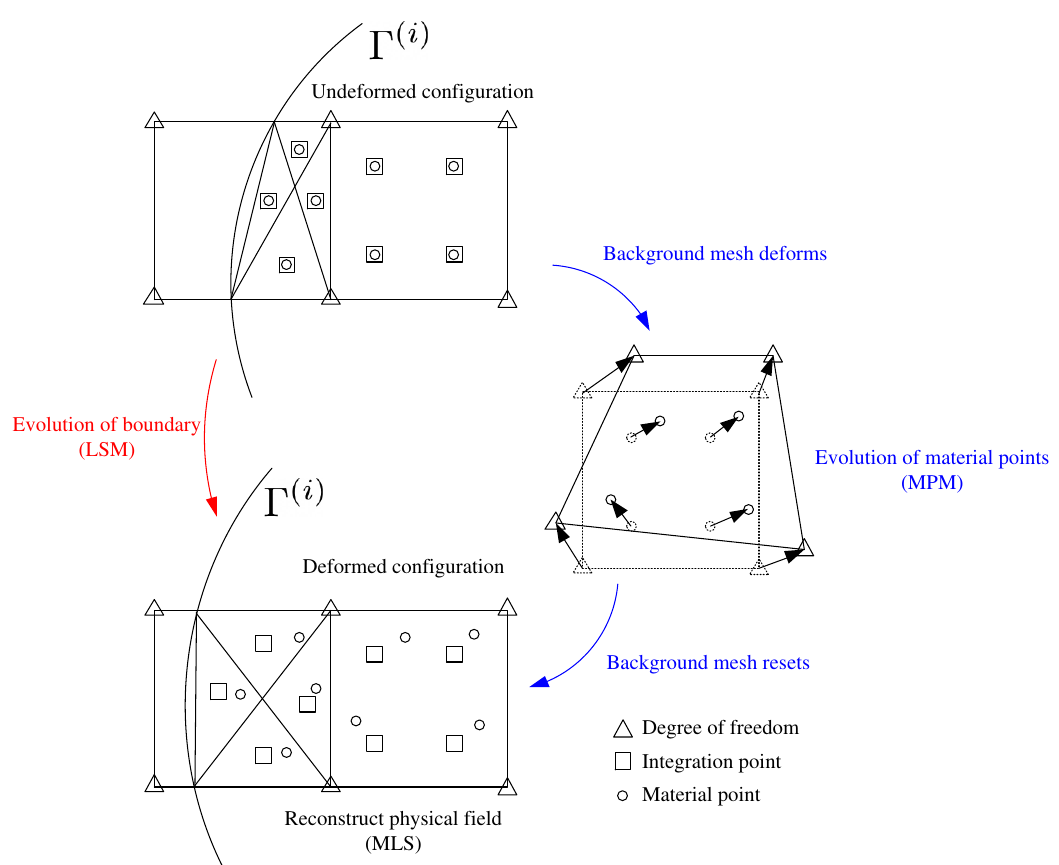}
\caption{Spatial discretization and information update.}
\label{figure_mpm}
\end{figure}

We now consider the discretization and update of $\Omega^{(i)}$. As shown in Fig. \ref{figure_mpm}, the unknowns are set at the nodes belonging to the elements inside of $\Omega^{(i)}$ or cut by the boundary $\Gamma^{(i)}$, e.g. the triangle labels in Fig. \ref{figure_mpm}. The location of $\Gamma^{(i)}$ is implicitly represented by the level set constructed at the Eulerian grid (consistent with the background grid of the MPM). For the element inside of $\Omega^{(i)}$, the locations of the integration points are identical to the typical Gauss's quadrature scheme. For the element cut by the boundary, we partition the element into a set of sub-triangles in terms of the location of $\Gamma^{(i)}$ (the boundary is assumed as a straight line in the element) and the integration points are at the centers of the sub-triangles. This integration scheme based on the element-partitioning is developed from the X-FEM \citep{sukumar2000extended}. Recently, an integration scheme without element-partitioning is also developed, see \cite{chin2015numerical, chin2017modeling, liu2019modeling2}. The integration points are labeled as squares in Fig. \ref{figure_mpm}. Initially, the material points share the same positions of the integration points (as shown in the undeformed configuration in Fig. \ref{figure_mpm}). The mesh then deforms to achieve an equilibrium state and the movements of the nodes are projected to the material points via the weighting functions, as shown in the vertically middle inset in Fig. \ref{figure_mpm}. In the next time step, the physical variables of the material points, such as location and stress, are updated and the background mesh is reset. Meanwhile, the level set evolves to update the boundary. In the typical implicit MPM, the elements containing material points are activated for the next calculation and the material points are always working as integration points, where the volume integration is inaccurate. In the latest MPM papers \citep{gong2015improving, sulsky2016improving}, the integration points are set in a same procedure described above considering the updated location of the boundary (as shown in the deformed configuration in Fig. \ref{figure_mpm}). Therefore, it requires to recover the physical variables at the integration points from the scattered material points and the MLS method meets the requirement.

In the MLS method, the reconstruction function of a scalar field $u^h(x)$ in one-dimension is assumed as holding the following form:
\begin{equation}
\label{eq:uh}
u^h(x,\bar{x}) = a_0(\bar{x}) + a_1(\bar{x}) x + \cdots + a_r(\bar{x}) x^r = \vec{P}^T(x) \vec{a}(\bar{x}),
\end{equation}
where the coefficients of the monomials $\vec{P}^T(x) = [1,x,x^2,\cdots, x^r]$, depend on the spatial coordinate $\bar{x}$, i.e. $\vec{a} = \vec{a}(\bar{x})$, $r$ is the the maximum degree of the monomials, a bar over $x$ for $\vec{a}$ is adopted to distinguish $x$ for $\vec{P}$. The general idea of the MLS method is to find $\vec{a}(\bar{x})$ so that the approximation function $u^h(x,\bar{x})$ minimizes the following functional:
\begin{equation}
\label{eq:ia}
I(\vec{a}) = \sum_{p=1}^{np} w(\bar{x}-x_p) \left[u^h_p(x_p,\bar{x}) - u_p\right]^2,
\end{equation}
where $p$ represents the $p$-th particle in $np$ particles, $w(\bar{x}-x_p)$ is the weighting function of particle $p$, $u_p = u(x_p)$ is the known value at particle $p$. By replacing the $\bar{x}$ in (\ref{eq:ia}) with $x$ and substituting (\ref{eq:uh}) into (\ref{eq:ia}), we obtain that:
\begin{equation}
I(\vec{a}) = \sum_{p=1}^{np} w(x-x_p) \left[\vec{P}^T(x_p) \vec{a}(x) - u_p\right]^2.
\end{equation}

Take the derivative of $I$ with respect to each component of $\vec{a}$ and set them to zeros to get:
\begin{equation}
\tensor{M}(x) \vec{a}(x) = \vec{b}(x),
\end{equation}
where $\tensor{M}(x)$ and $\tensor{b}(x)$ are defined as follows:
\begin{gather}
\tensor{M}(x) = \sum_{p=1}^{np} w(x-x_p) \vec{P}(x_p) \vec{P}^T(x_p),\\
\tensor{b}(x) = \sum_{p=1}^{np} w(x-x_p) \vec{P}(x_p) u_p.
\end{gather}
The approximation of $u(x)$ thus can be:
\begin{equation}
\label{eq:projection}
u^h(x) = \vec{P}^T(x) \tensor{M}^{-1}(x)\tensor{b}(x).
\end{equation}

For two-dimensional problems, we employ linear polynomials as the basis for the reconstruction, i.e.  $\vec{P}^T(\vec{x}) = [1,x_{1}, x_{2}]$ and a quadratic spline as the weighting function:
\begin{equation}
w(\vec{x} - \vec{x}_{p}) = \left\{
\begin{aligned}
&\frac{3}{4} - r^2 \quad & r \leq \frac{1}{2} \\
&\frac{1}{2} \left(\frac{3}{2} - r\right)^2 \quad & \frac{1}{2} \leq r \leq \frac{3}{2} \\
&0 \quad & r \geq \frac{3}{2}
\end{aligned}
\right.,
\end{equation}
where $r = ||\vec{x} - \vec{x_{p}}||/h$, $|| \cdot ||$ denotes the $L_{2}$ norm and $h$ is the grid spacing.
For convenience, we incorporate the open-source numerical analysis and data processing library (ALGLIB) \citep{ALGLIB} into our code for the interpolations, where the $k$-d tree is adopted for the efficient neighbor searching.

\subsection{Convergence scheme}
In this work, we here consider linear elastic materials, so the tangent matrix for the term in (\ref{eq:weak}) involving volume integration is conventional. We here only show the tangent for the terms involving integrals along $\Gamma_c$ and the finite difference method is employed to compute the Jacobian determinant matrix \citep{prevost2016faults}.

\begin{figure}[h]
\centering
\includegraphics[width=6cm]{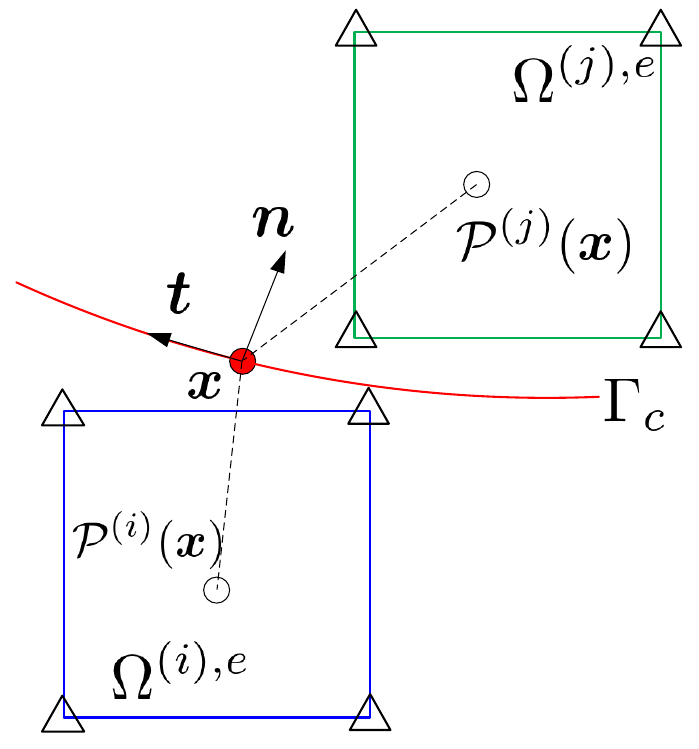}
\caption{Computation of tangent matrix for surface integrations}
\label{figure_twoelements}
\end{figure}

From (\ref{eq:weak}), the residual contributed from the surface integral is termed as contact residual and defined as:
\begin{equation}
r_c (\vec{u}, \vec{v}) := \int_{\Gamma_{c}} \llbracket \vec{v}\rrbracket^{(ij)}_{\rm{int}} \vec{\tau}(\vec{u}) d\gamma =
\int_{\Gamma_c} j^{(ij)}_{\rm{int}} (\vec{v}) \tau_n(\vec{u}) d\gamma  +
\int_{\Gamma_{c}} s^{(ij)}_{\rm{int}} (\vec{v}) \tau_{t}(\vec{u}) d\gamma,
\end{equation}
where $\llbracket \vec{v}\rrbracket^{(ij)}_{\rm{int}} = \vec{n} j^{(ij)}_{\rm{int}} (\vec{v})  + \vec{t} s^{(ij)}_{\rm{int}} (\vec{v}) $ is the displacement jump and $\vec{\tau}(\vec{u}) = \vec{n}  \tau_n(\vec{u}) + \vec{t}  \tau_{t}(\vec{u})$ is the traction along the contact force. As shown in Fig. \ref{figure_twoelements}, the two projected points $\mathcal{P}^{(i)}(\vec{x})$ and $\mathcal{P}^{(j)}(\vec{x})$ of the surface integration point $\vec{x} \in \Gamma_c$ are within elements $\Omega^{(i),e}$ and $\Omega^{(j),e}$, respectively. We construct a generalized nodal displacement vector $\{\vec{u}\} = \{\vec{u}^{(i),e}, \vec{u}^{(j),e}\}$ and a shape function matrix $[\tensor{N}] = [-\tensor{N}^{(i)}(\mathcal{P}^{(i)}(\vec{x})), \tensor{N}^{(j)}(\mathcal{P}^{(j)}(\vec{x}))] $, where $\vec{u}^{(i),e}$ and $\vec{u}^{(j),e}$ are nodal displacement of elements $\Omega^{(i),e}$ and $\Omega^{(j),e}$, respectively, and $\tensor{N}^{(i)}(\mathcal{P}^{(i)}(\vec{x}))$ and $\tensor{N}^{(j)}(\mathcal{P}^{(j)}(\vec{x}))$ are nodal shape functions at the projected points. We then implement a radial extension of displacement \citep{osher2006level} leading to $\vec{u}^{(i)}(\vec{x}) = \vec{u}^{(i)}(\mathcal{P}^{(i)}(\vec{x}))$ and $\vec{u}^{(j)}(\vec{x}) = \vec{u}^{(j)}(\mathcal{P}^{(j)}(\vec{x}))$. Therefore, the virtual displacement jump can be expressed as:
\begin{equation}
\llbracket \vec{v}\rrbracket^{(ij)}_{\rm{int}} = [\tensor{N}]^T\{\vec{v}\},
\end{equation}
where the vector $\{\vec{v}\}$ is the general virtual displacement vector defined in a same way of $\{\vec{u}\}$. Due to the arbitrariness of $\vec{v}$, the contribution of $\vec{x}$ to the residual of $\vec{r}_c (\vec{u}, \vec{v})$ in a matrix form is:
\begin{equation}
\label{eq:rc}
\{\vec{r}_c\} = \int_{\Gamma_{c}} [\tensor{N}]^T
\renewcommand\arraystretch{0.7}
\begin{bmatrix}
n_x & t_x\\
n_y & t_y \\
\end{bmatrix}
\setlength\arraycolsep{2pt}
\left\{
\begin{array}{c}
\tau_n \\
\tau_t
\end{array}
\right\}
d\gamma.
\end{equation}

Since the tangential traction would be affected by the normal traction, we here adopt the finite-differencing of the residual to compute the Jacobian determinant rather than deriving theoretical tangent matrix for simplicity. The elemental contribution to the Jacobian determinant matrix for the contact residual is:
\begin{equation}
\label{eq:jpq}
[J^e_{PQ}] = \frac{\partial r_P}{\partial u_Q} \approx \frac{r_P(u_Q + \bar{h}) - r_P(u_Q)}{\bar{h}},
\end{equation}
where $P$ and $Q$ are global equation numbers corresponding to unknown displacements. The perturbation of the degrees of freedom $\bar{h}$ is a small parameter chosen as:
\begin{equation}
\bar{h} = \frac{\sqrt[3]{\epsilon_M}}{|u_Q|}, \quad \bar{h} = \max(\sqrt[3]{\epsilon_M}, \bar{h}),
\end{equation}
where $\epsilon_M$ is machine precision.

Once we achieve the elemental contribution to the Jacobian determinant matrix from the contact residual, we can easily obtain the global tangent matrix via assembling Jacobian determinant matrix  element-by-element and the regular stiffness matrix. We next discuss the calculation of the traction along the contact surface. Again, the traction is updated by applying the classical return mapping algorithm over the incremental form of (\ref{eq:yield}). We here modify the algorithm of \cite{annavarapu2014nitsche}, which is shown in Algorithm \ref{algorithm:columb}.

\begin{algorithm}
\caption{Update tractions on the contact surface at the $(k + 1)$-th iteration given converged results at $k$-th iteration}
\begin{algorithmic}
\For {all active Gauss-points on $\Gamma_c$}

\State Compute normal gap $g_{n}^{(k+1)}$ as (\ref{eq:gn})

\State Compute tangential gap $g_t^{(k+1)}$ as (\ref{eq:s})

\State Compute trial normal traction $\tau^{tri, (k+1)}_{n}$ as (\ref{eq:tau_n})

\State Compute trial tangential traction $\tau^{tri,(k+1)}_{t} = \frac{1}{\epsilon_t} (g_t^{(k+1)} - g_t^{p, (k)})$

\State Compute trial yield function $\phi^{tri, (k+1)} = \phi(\tau^{tri,(k+1)}_{t}) = |\tau^{tri,(k+1)}_{t}| - \mu |\tau^{tri,(k+1)}_{n}|$

\If {$\phi^{tri, (k+1)} \leq 0$}
    \State Trial state is true state: $\tau^{(k+1)}_{n} = \tau^{tri, (k+1)}_{n}$ and $\tau^{(k+1)}_{t} = \tau^{tri, (k+1)}_{t}$
\Else
    \State Normal direction - no yielding: $\tau^{(k+1)}_{n} = \tau^{tri, (k+1)}_{n}$
    \State Tangential direction - return-mapping algorithm
    \State $\Delta \gamma = \epsilon_t \phi^{tri, k}$
    \State $\tau^{k}_{t} = \tau^{tri, k}_{t} - \frac{\Delta \gamma}{\epsilon_t} \frac{\tau^{tri, k}_{t}}{|\tau^{tri, k}_{t}|}$
    \State $g_t^{p, (k+1)} = g_t^{p, (k)} + \Delta \gamma \frac{\tau^{tri, k}_{t}}{|\tau^{tri, k}_{t}|}$
\EndIf
\EndFor
\end{algorithmic}
\label{algorithm:columb}
\end{algorithm}

\subsection{Calculation procedure}
The calculation procedure is detailed as follows.
\begin{enumerate}
\item Choose a proper spatial region $\mathcal{B}$ containing all bodies.
\item Initialize the level set for each body $\Phi^{(i)}(\vec{x})$, $\forall i\in \mathcal{I}$.
\item Generate a set of material points for each body according to its boundary $\Gamma^{(i)}$ and set unknowns regarding to the boundary conditions, as shown in Fig. \ref{figure_mpm}.
\item Determine the potential contact pairs and set the reference contact surface $\Gamma_c$ as (\ref{eq:gammac}).
\item Implement the Newton-Raphson iteration scheme to solve the displacement vanishing the residuals shown in (\ref{eq:weak}) by repeating the following steps ($k$ representing the $k$-th iteration).
\begin{enumerate}
\item Compute the tractions for all integration points $\vec{x}_p$ on the $\Gamma_c$, according to Algorithm \ref{algorithm:columb}.
\item If $\tau_n(\vec{x}_p) \leq 0$ at $\vec{x}_p$, compute the contribution to the contact residual according to (\ref{eq:rc}) and its contribution to Jacobian matrix according to (\ref{eq:jpq}).
\item Considering the stiffness matrix coming from the traditional volume integration, assemble the global tangent matrix $\tensor{T}^{(k)}$.
\item Compute the global residual $\vec{r}^{(k)}$.
\item Update the displacement by $\vec{u}^{(k+1)}=\vec{u}^{(k)} + \delta \vec{u}^{(k)}$, where $\delta \vec{u}^{(k)} = (\tensor{T}^k)^{-1} \vec{r}^{(k)}$.
\end{enumerate}
\item Update the locations and stresses of the materials points, as shown in Fig. \ref{figure_mpm}.
\item Extend the displacement of each body to the whole region $\mathcal{B}$ according to the steady state of (\ref{eq:extension}).
\item Evolve the level set $\Phi^{(i)}(\vec{x})$, $\forall i\in \mathcal{I}$, according to (\ref{eq:phit}).
\item Set the integration points and unknowns according to the updated level sets and boundary conditions.
\item Project the information of the scattered material points to the integration points, according to (\ref{eq:projection}).
\item Repeat from step 4.
\end{enumerate}

One last remaining problem is the setting of the penalty parameters $\epsilon_n$ in (\ref{eq:tau_n}) and $\epsilon_t$ in (\ref{eq:tau_t}). Following the appoarch from \cite{kikuchi1988contact}, which is also adopted in \cite{leichner2019contact}, we set the parameters in relation to the average Young's modulus $\bar{E}$ and mesh size $h$:
\begin{equation}
\epsilon_n = \epsilon_t = \epsilon_0 \frac{h}{\bar{E}}.
\end{equation}
We hereafter set $\epsilon_0 = 1$ for all cases in terms of accuracy and convergence rate. Furthermore, this factor is kept as a constant during the iterations.

\section{Numerical Examples}
In this section, we provide numerical examples to verify the model and demonstrate capacities. We first consider the cases containing a single contact to verify the implementation via comparisons of our results and results in literature or analytical solutions. The contacts involving multiple bodies with simple and complicated shapes are then simulated to demonstrate the efficiency of the proposed models to capture complex contacts. The last two examples are with multiple loading steps considering the evolution of boundaries, and one is used to validate and the other to show the prospective of our model in terms of linking particle physical states and macroscopic responses of the assemblies.


\subsection{Verification Problems}
\subsubsection{Two contacting blocks}
\begin{figure}[h]
\centering
\includegraphics[width=7cm]{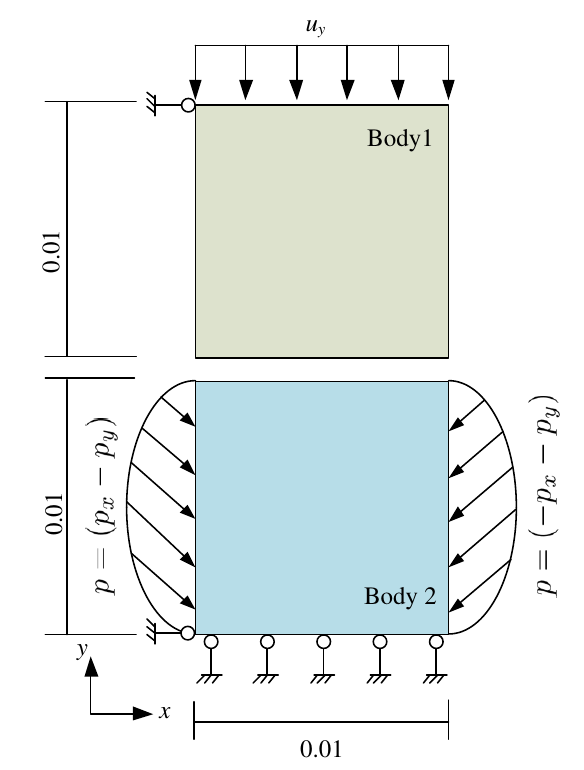}
\caption{Model and boundary conditions}
\label{case1_model}
\end{figure}

To verify our implementation, we first simulate the deformation of two contacting rectangular blocks discretized by structured cells, as discussed in \cite{tur2009mortar}. As shown in Fig. \ref{case1_model}, a vertical displacement $u_y=-1.6\times10^{-6}$ m is imposed on the top boundary of the upper body and stress distributions $p_x = 4 \times 10^{11} y (0.01-y)$ Pa and $p_y = 10^{12}y(0.01- y)$ Pa are applied on the left and right sides of the lower body. The origin of the coordinate system locates at the lower left corner of the lower body. Other boundary conditions and geometries are shown in Fig. \ref{case1_model}. Linear elastic material is assumed with plane strain and Young's modulus $E=100$GPa and Possion's ratio $\nu = 0.3$ for the two bodies. Coulomb model with coefficient of friction $\mu = 1.0$ is implemented to distinguish stick and slip regions. Contact reference is exactly set along the line $y=0.01$ m and the mesh size is set as $h=0.0002$ m.

\begin{figure}[h]
\centering
\includegraphics[width=11cm]{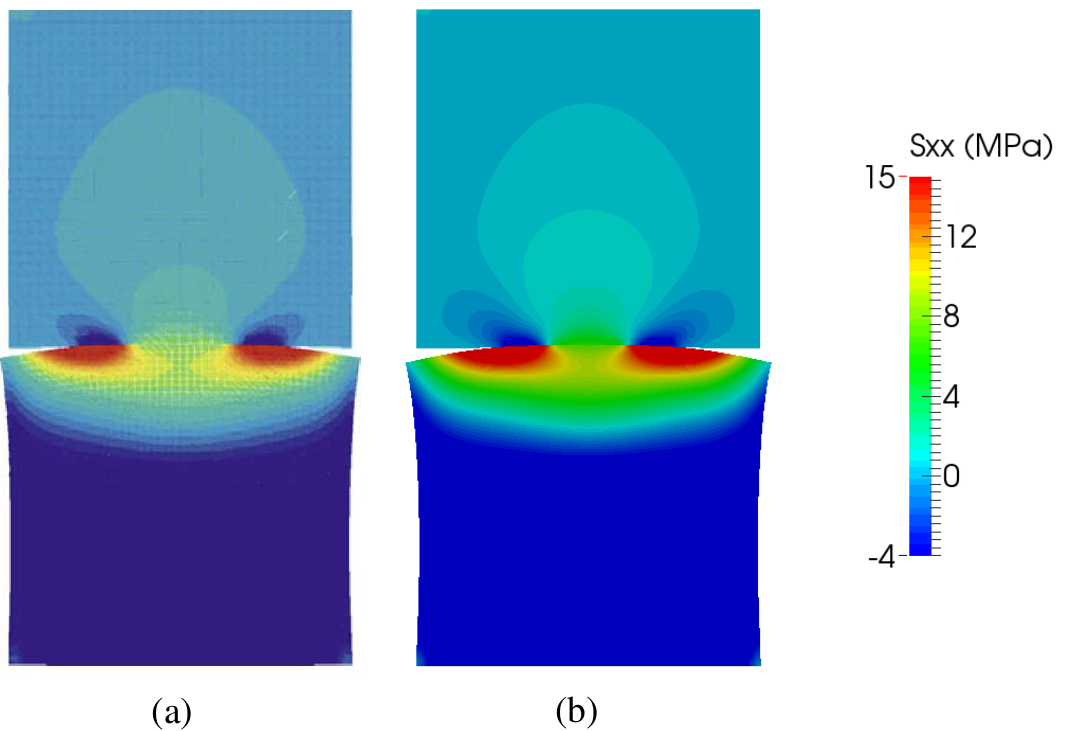}
\caption{Comparison of stress component $\sigma_{1}$: (a) results in \cite{tur2009mortar} and (b) our results.}
\label{case1_sxx}
\end{figure}

\begin{figure}{h}
\centering
\includegraphics[width=7cm]{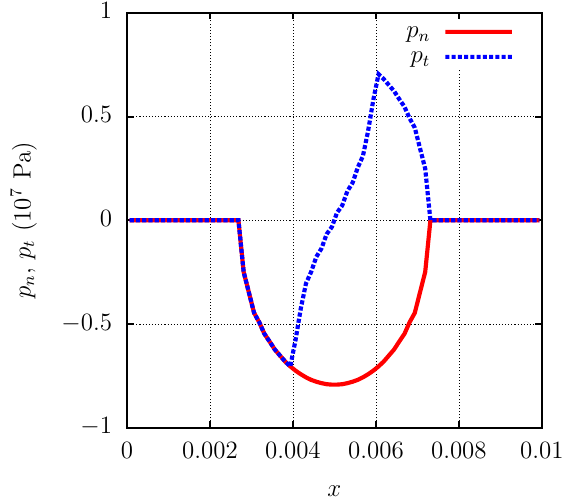}
\captionof{figure}{Contact stresses: $p_n$ is the normal traction and $p_t$ is the tangential traction.}
\label{fg:case1_pnt}
\end{figure}

Fig. \ref{case1_sxx} shows the comparison of our results for distribution of $\sigma_x$ on the distorted domain (amplified by a factor of 500) and the results shown in \cite{tur2009mortar}. We can see that the distributions are identical. Fig. \ref{fg:case1_pnt} shows the contact tractions along the contact surface. As can be seen, the contact area is split into a central stick zone and two slip zones with opposite slip directions, which also coincides with the observation in \cite{tur2009mortar}. Unlike setting two loops to vanish the residuals and to ensure the correct shear states \citep{tur2009mortar}, we here do not need to distinguish the shear states since the Coulomb friction model is treated as an elastoplastic model and a typical return-mapping algorithm is implemented to converge the simulation. The convergence for this example is achieved within eight iterations.

\subsubsection{Two contacting trapezoids}
\begin{figure}[h]
\centering
\includegraphics[width=11cm]{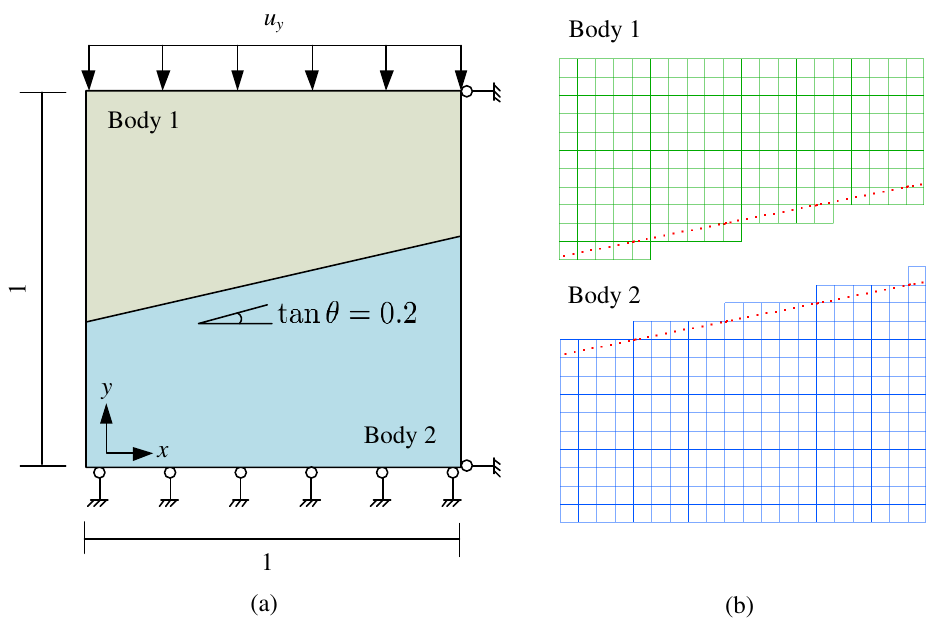}
\caption{Two contacting bodies with an inclined contact plane (a) loading and boundary conditions and (b) meshes used in computations (red dots denoting the contact monitoring points on the potential contact surface).}
\label{case2_model}
\end{figure}

In this example, we consider two bodies contacting with each other along an inclined plane as proposed in \cite{annavarapu2014nitsche}. As shown in Fig. \ref{case2_model}a, the equation of the inclined plane is $y-0.2x - 0.4586=0$ leading a slope with $\tan \theta = 0.2$. A vertical displacement $u_y=-10^{-3}$ m is imposed on the top boundary of the upper body. Other boundary conditions and geometries can be found in Fig. \ref{case2_model}a. Young's modulus and Possion's ratio are $E=1$GPa and  $\nu = 0.3$, respectively, for the two bodies. Fig. \ref{case2_model}b shows the meshes used in computations and the red dots denote the contact monitoring points on the contact reference. Note that we here separately treat each body holding its own meshes rather than enrich DOFs for only one grid as the X-FEM.

\begin{figure}[h]
\centering
\includegraphics[width=11cm]{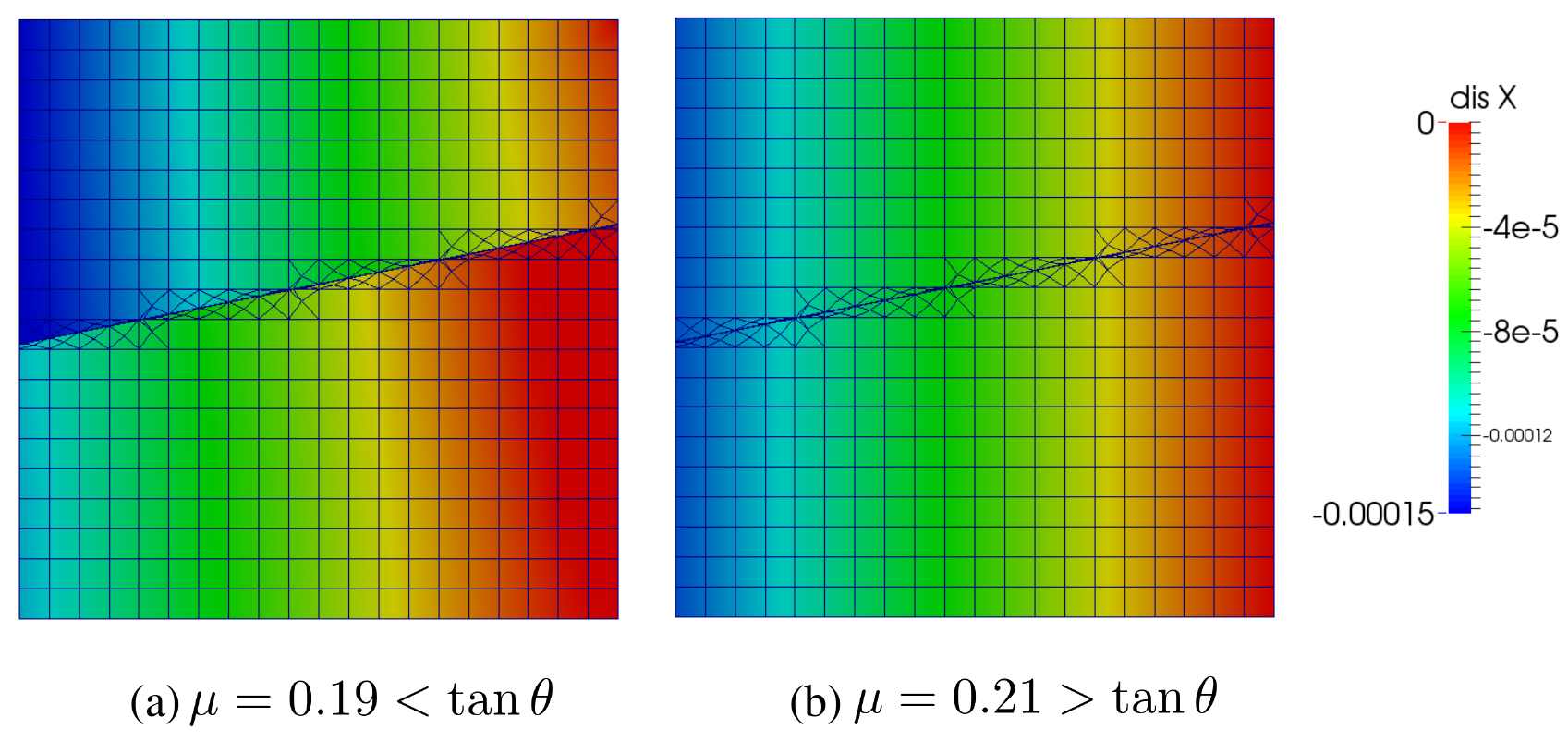}
\caption{Horizontal displacement contours for different friction coefficient (a) friction angle is less than incline angle with slip expected and (b) friction angle is greater than incline angle with stick expected.}
\label{case2_result}
\end{figure}

The friction coefficient is chosen as $\mu = 0.19$ and $\mu=0.21$ for two separate computations. The problem serves as a good benchmark for fictional sliding problems since we can predict slipping behavior when the friction coefficient $\mu < \tan \theta$ and a stick state otherwise. Fig. \ref{case2_result} shows the distributions of horizontal displacements for different friction coefficients. As expected, slipping occurs when the friction coefficient is less than the tangent of the inclination of the surface while it is larger, we see a stick state. We repeat that the cells used for computations are squares as shown in Fig. \ref{case2_model}b and the subcells (triangles) partitioning the squares cut by the contact surface are only used for volume integrals and post-processing as shown in Fig. \ref{case2_result}. Hereafter, all results containing cells are plotted following this treatment except for some specifications.

\subsubsection{Hertzian contact}
\begin{figure}[h]
\centering
\includegraphics[width=12cm]{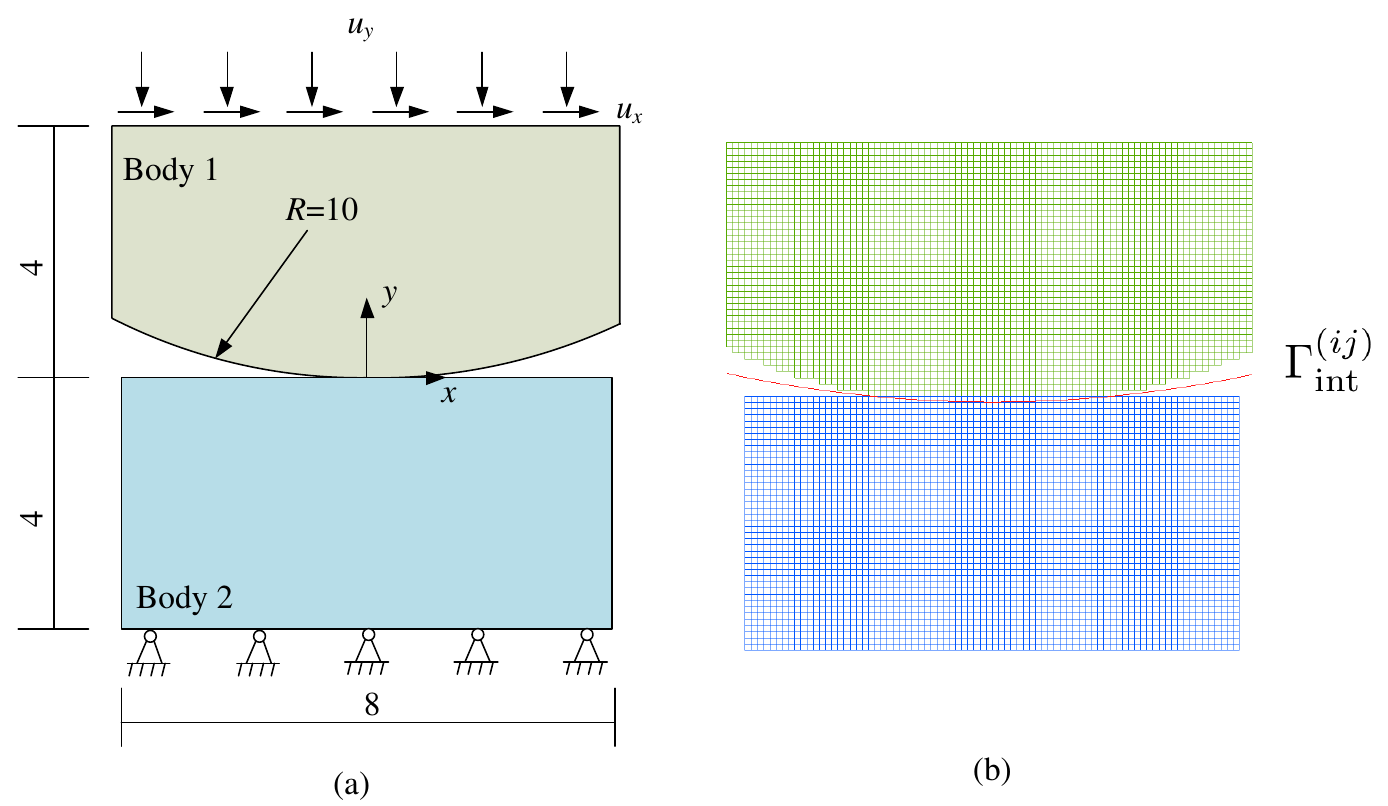}
\caption{Hertz contact problem (a) geometry and boundary conditions (b) meshes and intermediate contact reference.}
\label{case3_model}
\end{figure}

The third problem to validate our model is the cylinder one plane Hertzian contact. It has a closed analytical solution for the contact stresses and so is used as a comparison with the numerical results. Fig. \ref{case3_model}a shows the boundary conditions and geometries. The radius of the cylinder is 10 mm and the height of each body is 4 mm.  Young's modulus is 10 GPa and Poisson's ratio is 0.3 for both bodies. A displacement with $\vec{u} = (0.03, -0.04)$ mm is applied on the top surface of the cylinder part. Fig. \ref{case3_model}b shows the meshes used in the computations and the contact reference. The mesh size is $h=0.1$ mm.

\begin{figure}[h]
\centering
\includegraphics[width=10cm]{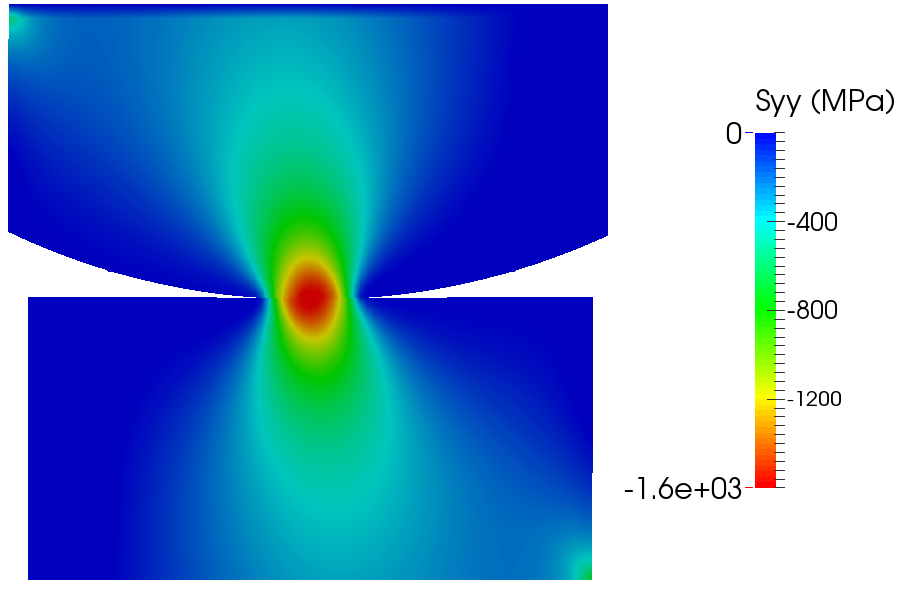}
\caption{Distribution of vertical stress.}
\label{case3_syy}
\end{figure}

\begin{figure}[h]
\centering
\includegraphics[width=11cm]{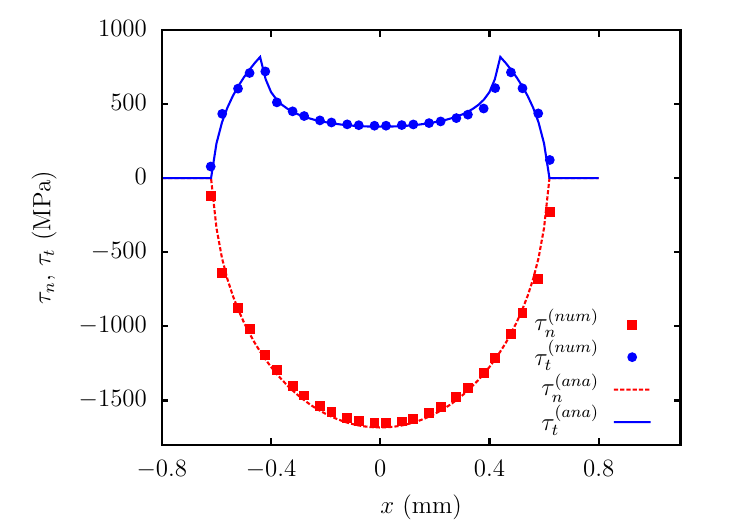}
\caption{Comparison of numerical results and analytical results for tractions along the contact surface (subscripts "num" denoting numerical results and "ana" representing analytical results).}
\label{case3_comparison}
\end{figure}

Fig. \ref{case3_syy} plots the distribution of the vertical stress. In order to compare the numerical and analytical results, we first compute the reaction forces acting on the top boundary, as $\vec{f} = (-570.73, 1617.64)$ kN. The distributions of tangential and normal tractions obtained from simulations are compared with the analytical solutions as shown in Fig. \ref{case3_comparison}. We can clearly see a stick zone existing between two slip zones and distributions of tractions are almost identical.

\subsection{Frictional contacts involving multiple bodies}
\subsubsection{Symmetric problems for nine-discs under an isotropic compression}
\begin{figure}[h]
\centering
\includegraphics[width=11cm]{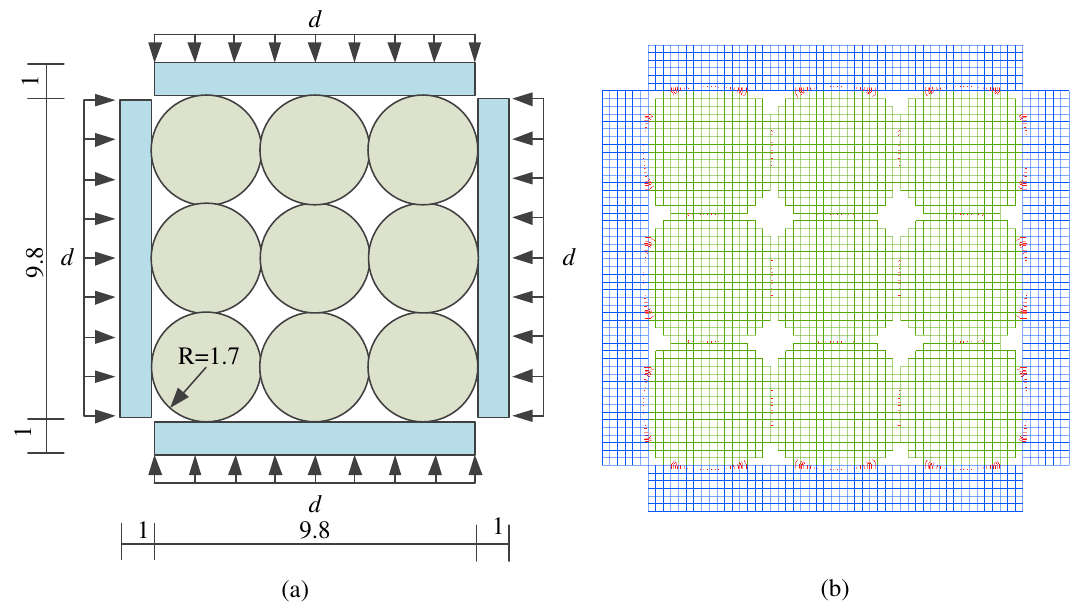}
\caption{Problem for nine particles: (a) geometry and boundary conditions and (b) meshes and closest projections from contact reference to boundaries.}
\label{case4_model_mesh}
\end{figure}

We here further verify our model for symmetrical problems. As shown in Fig. \ref{case4_model_mesh}a, nine discs with same radius of 1.7mm are confined by four deformable plates. The thickness of the confining plates is 1mm and the length 9.8mm. An isotropic compression is imposed by moving the plates towards the center with a displacement $d=10^{-4}$mm. Note that even we only fix the confining plates without any displacement, the discs are compressed since there are initial gaps between bodies. Fig. \ref{case4_model_mesh}b shows the meshes used for the computations. The material is also assumed as linear elastic with Young's modulus $E=100$ GPa and Poisson's ratio $\nu=0.3$. The friction coefficient is set as 0.5.

\begin{figure}[h]
\centering
\includegraphics[width=15cm]{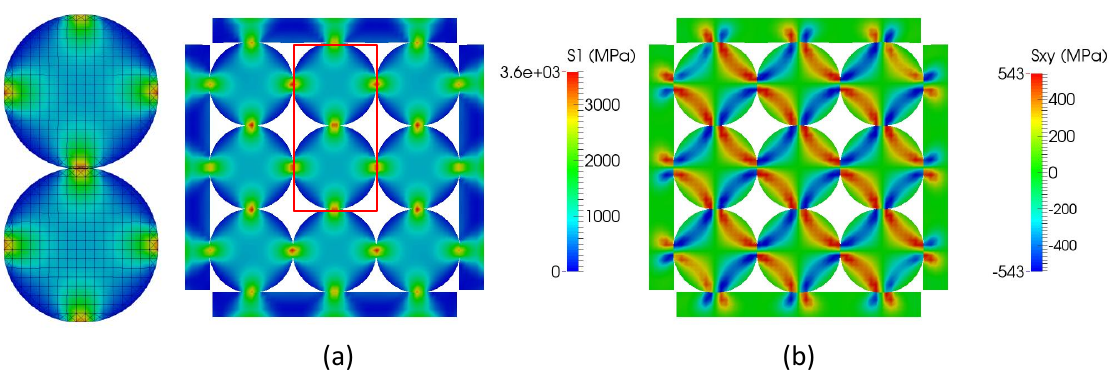}
\caption{Distributions of stresses (a) the maximum principle stress and (b) shear stress.}
\label{case4_results}
\end{figure}

The computation rapidly converged within four iterations due to the high symmetry of the problem. Fig. \ref{case4_results} shows the distributions of the maximum principle stress and the shear stress. The left inset details the meshes for two contacting bodies, where we can find that high stress concentrations occur at the contacting points. In a summary, all stresses are symmetrically distributed consistent with expectations.

\subsubsection{Brazilian test for a grain obtained from Micro-CT images}
\begin{figure}[h]
\centering
\includegraphics[width=15cm]{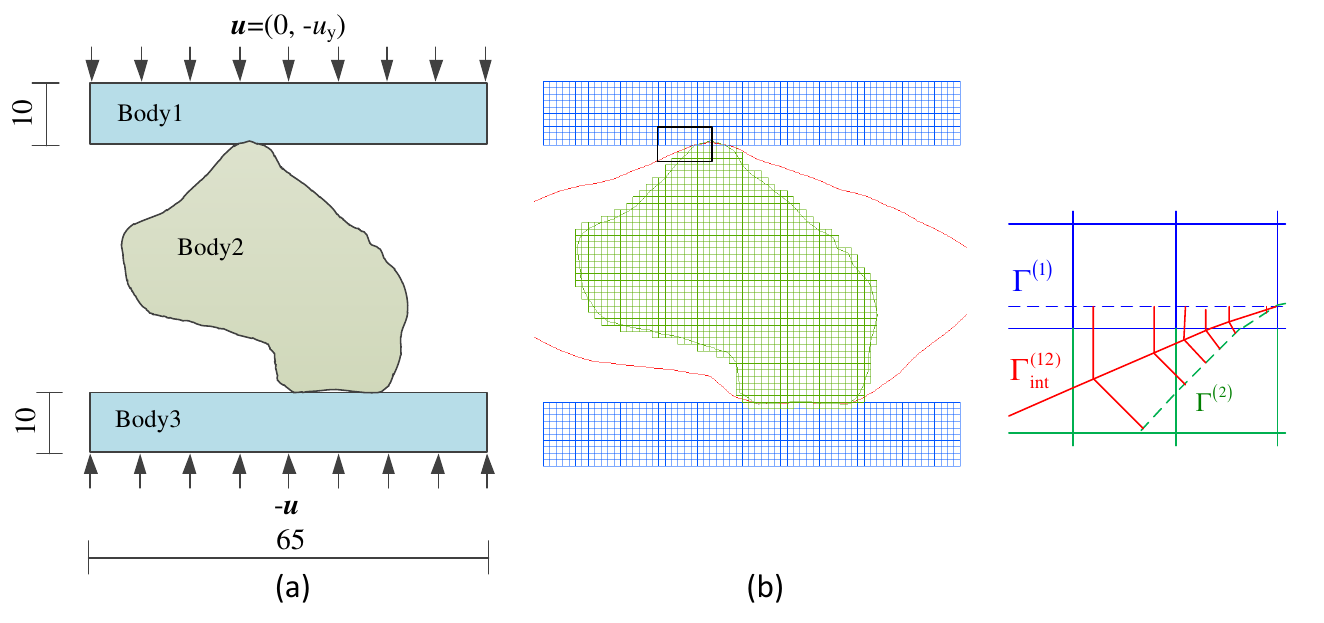}
\caption{Brazilian test for a real sand based on Micro-CT scanning images: (a) geometry and boundary conditions and (b) meshes and contact references (inset: local enlargement of closest projections).}
\label{case4b_model_mesh}
\end{figure}

In this example, we simulate the Brazilian test for a grain whose geometry is inferred from micro-CT scanning images. Since we only consider two-dimensional problems in this paper, the geometry of the grain is a cross section of a real Hostun grain reconstructed based on a binarized 3D image \citep{gupta2019open}, which is also used in \cite{liu2019shift}. To increase the difficulty and test the robustness of the algorithm, we setup the configuration as shown in Fig. \ref{case4b_model_mesh}, where two potential contact points locate on the lower plate. The width and length of the confining plates are 10 mm and 65 mm, respectively. The displacement of the top boundary of the upper plate is fixed as $\vec{u} = (0, -0.1)$mm, and $\vec{u}=(0,0.1)$mm for the bottom boundary of the lower plate. Again, the materials for these three deformable bodies are identical and assumed as linear elastic with Young's modulus 10 GPa and Poisson's ratio 0.3. The friction coefficient is also set as 0.5. Fig. \ref{case4b_model_mesh}b shows contact references and the meshes used in the computation with the mesh size $h=1$ mm. The inset of Fig. \ref{case4b_model_mesh} illustrates the enlargement of the closest projections from the surface integration points on the contact reference surface to the boundaries of bodies.

\begin{figure}[h]
\centering
\includegraphics[width=15cm]{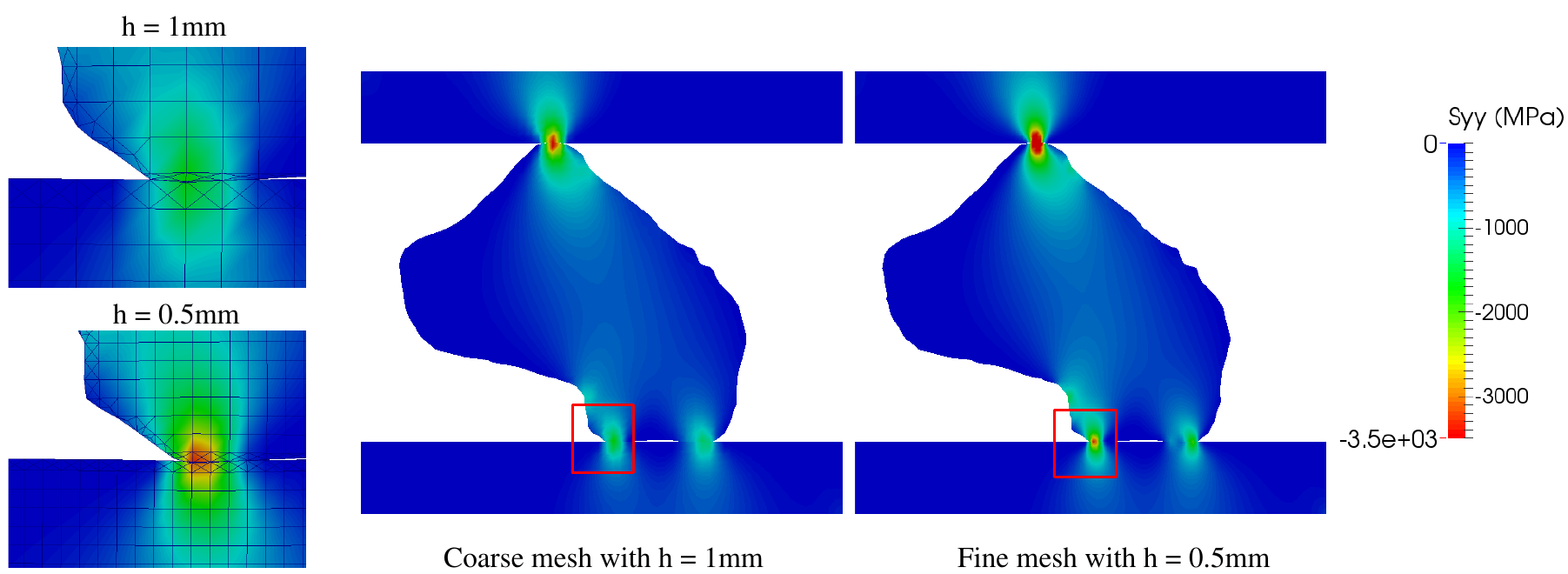}
\caption{Distributions of vertical stresses for different mesh sizes in the same legend.}
\label{case4b_results}
\end{figure}

We perform the simulations two times for different mesh sizes, i.e. $h=1$ mm and $h=0.5$ mm, to examine the sensitivity of result to the mesh size. Fig. \ref{case4b_results} compares the distributions of vertical stresses for these two mesh sizes with the same legend. The left inset of Fig. \ref{case4b_results} shows the enlarged contours at the left bottom contact points in different meshes. The contour patterns are consistent for various mesh sizes though the maximum values are a bit of different since the stress at the contacting point is highly concentrated. From this example, we can see the advantage of the proposed method over the DEM where we cannot simultaneously handle two contact points within one contact.

\subsubsection{Four real grains under an isotropic compression}
\begin{figure}[h]
\centering
\includegraphics[width=8cm]{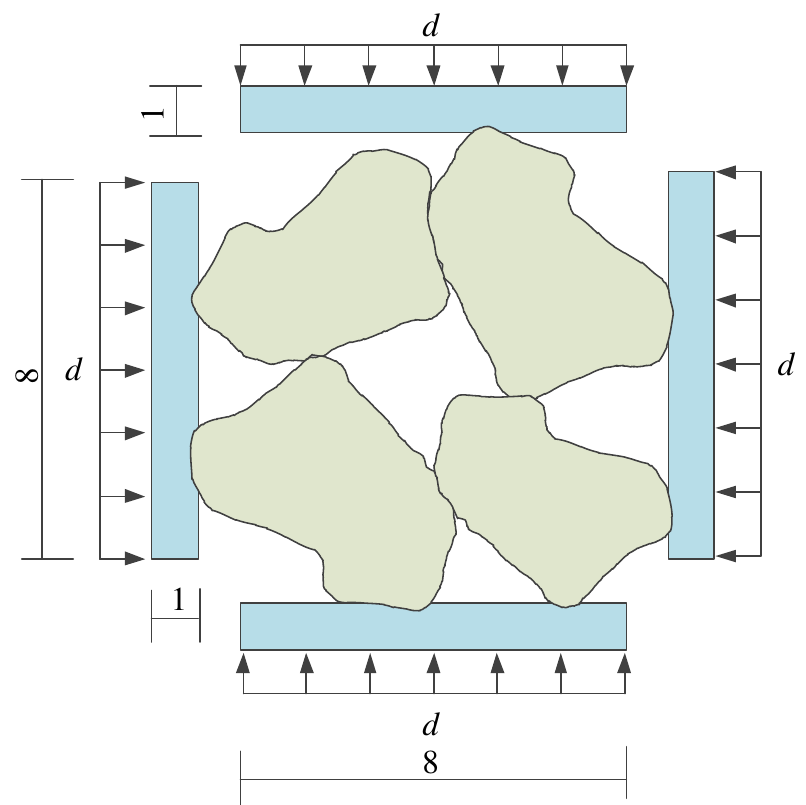}
\caption{Geometry and boundary conditions for the model of four grains under an isotropic compression.}
\label{case_four_grains_model}
\end{figure}

We deepen the last example to the case of four grains under an isotropic compression. The grains hold the same shape but with various sizes and locations via rotation and translation. Fig. \ref{case_four_grains_model} shows the initial configuration. Four deformable plates move toward the center with $d=10^{-3}$ mm. The material parameters are identical to the last example. The mesh size is set as $h=0.1$ mm. 

\begin{figure}[h]
\centering
\includegraphics[width=13cm]{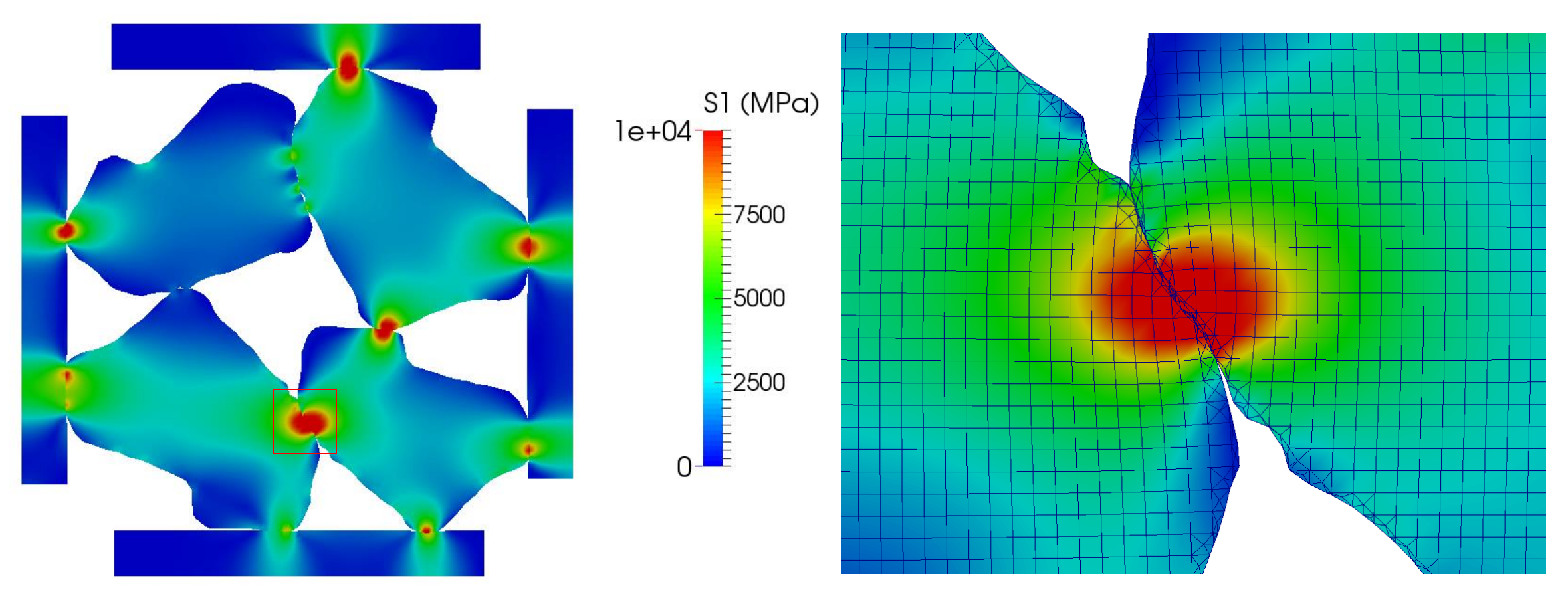}
\caption{Distribution of the maximum principle stress.}
\label{case_four_grains_results}
\end{figure}

Fig. \ref{case_four_grains_results} plots the distribution of the maximum principle stress and an enlargement for the vicinity of a specific contact point. We can observe an obvious shift between two stress concentration regions due to slipping.

\subsection{Contacts with incrementally increasing loads considering the evolution of level set}
In the following examples, we incrementally increasing the loading to consider the evolution of level set and material points. The first example to show the effectiveness of the proposed method to consider large deformation due to the essence of the MPM. The second example, we explore the capability of the proposed method to simulate an assembly of granular materials with more accurate information than the DEM.

\subsubsection{Brazilian test for a disc}
\begin{figure}[h]
\centering
\includegraphics[width=12cm]{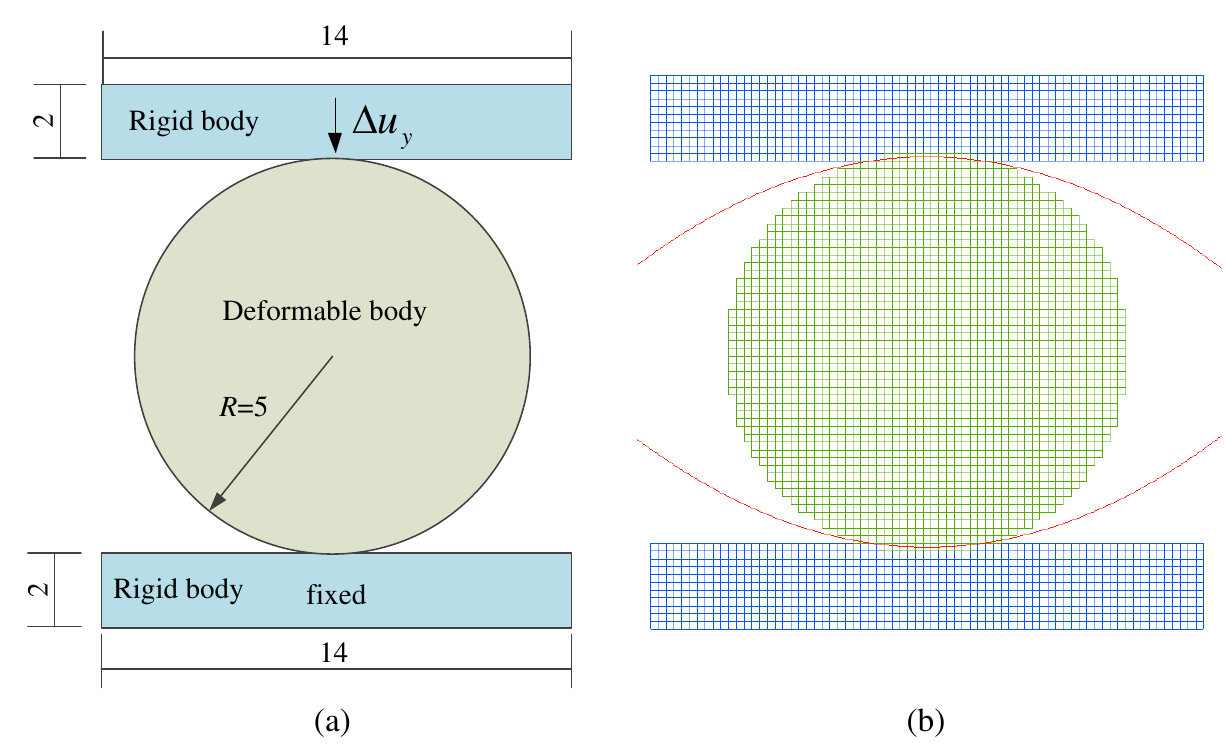}
\caption{Brazilian test for a disc: (a) geometry and boundary conditions and (b) meshes and contact references.}
\label{case_brazilian_disc_model}
\end{figure}

As shown in Fig. \ref{case_brazilian_disc_model}a, a deformable disc is vertically compressed by two rigid rectangles. The low rectangle is fixed and a constant vertical displacement $\Delta u_y = 0.02$ mm is applied on the top rectangle in each time step. The radius of the disc is $R=5$ mm. Other geometries can be found in Fig. \ref{case_brazilian_disc_model}a. The material for the disc is assumed as linear elastic with Young's modulus $E=100$ GPa and Poisson's ratio $\nu=0.3$. The friction coefficient is also set as 0.5. Note that the movements of the rectangles are prescribed as rigid bodies. Fig. \ref{case_brazilian_disc_model}b shows the contact references and the meshes for the initial configuration with mesh size $h=0.2$ mm. We simulate 50 steps.

\begin{figure}[H]
\centering
\includegraphics[width=9cm]{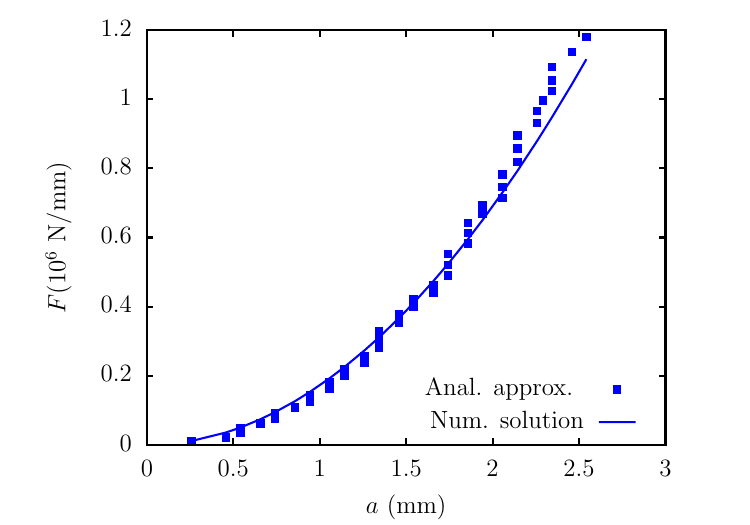}
\caption{Comparison of the relationships between contact radius and the contact force for the numerical and analytical solutions.}
\label{case_brazilian_disc_curve}
\end{figure}

Since the displacement of the top rigid rectangle is not equivalent to the penetration $d$ used in the Hertz theory, we validate our model by comparing the curves of contact radius vs. contact force, for the analytical and numerical solutions. According to the Hertz contact theory \citep{barber1992elasticity}, we have
\begin{equation}
\frac{a}{2} = \sqrt{\frac{F(1-\nu^2) R}{\pi E}},
\end{equation}
where $a$ is the radius of the contact surface and $F$ is the vertical contact force. For the top contact surface, we here approximate $a$ by the distance between the farthest activated integration point on the contact surface and the center of contact. Fig. \ref{case_brazilian_disc_curve} compares the relationship of $a-F$ for numerical and analytical solutions. The tendencies are identical although there are some discrepancies due to the inaccurate measurement of $a$.

\begin{figure}[H]
\centering
\includegraphics[width=15cm]{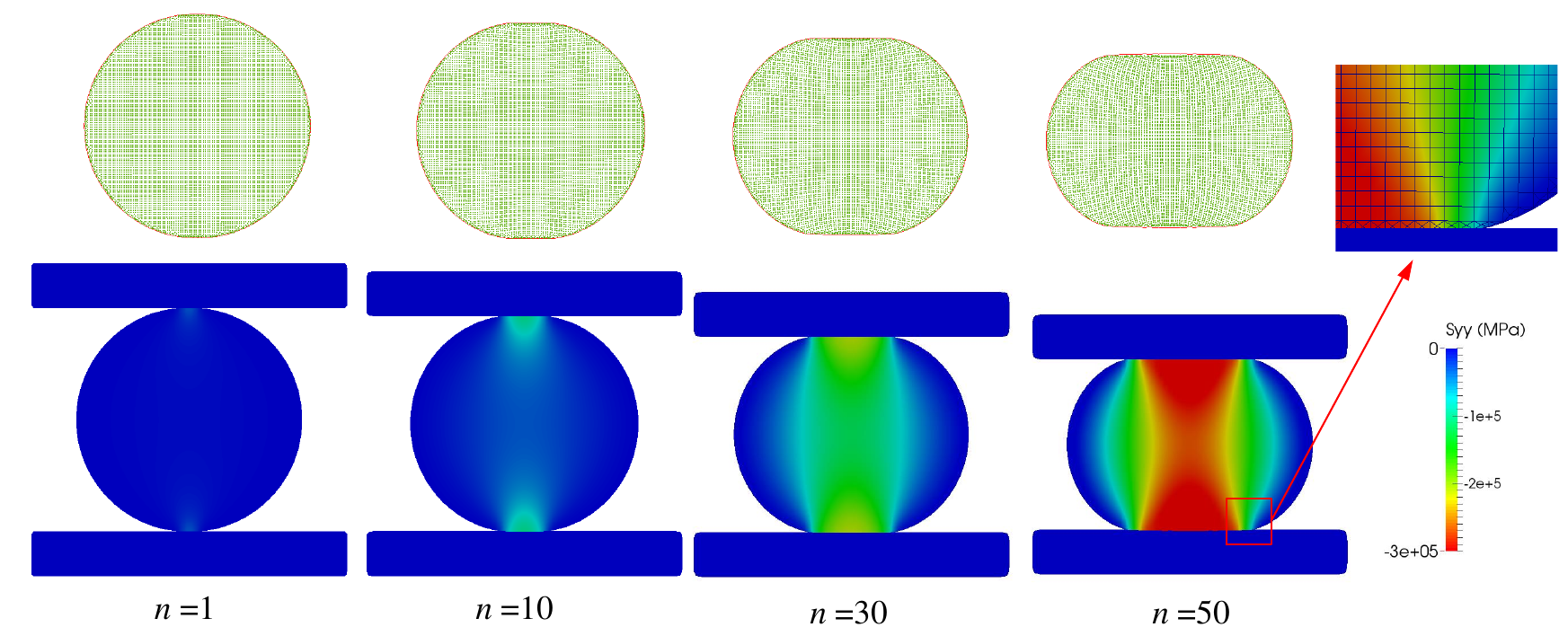}
\caption{Evolution of the vertical stress of the deformable disc (top row shows the locations of material points)}
\label{case_brazilian_disc_results}
\end{figure}

Fig. \ref{case_brazilian_disc_results} shows the evolution of vertical stress of the deformable disc at different loading steps. The top row of Fig. \ref{case_brazilian_disc_results} illustrates the locations of the material points. It should reminded that the material points carrying all physical information are not coincide with the integration points for the volume integrations. The elements used for the volume integrations (see the inset of Fig. \ref{case_brazilian_disc_results}) are regenerated according to the updated level set at the beginning of each time step. Therefore, we can avoid the issue of mesh distortion occurred in the FEM for such a problem and conserve the mass by tracking all material points.

\subsubsection{Fifteen non-sphere particles}
\begin{figure}[H]
\centering
\includegraphics[width=15cm]{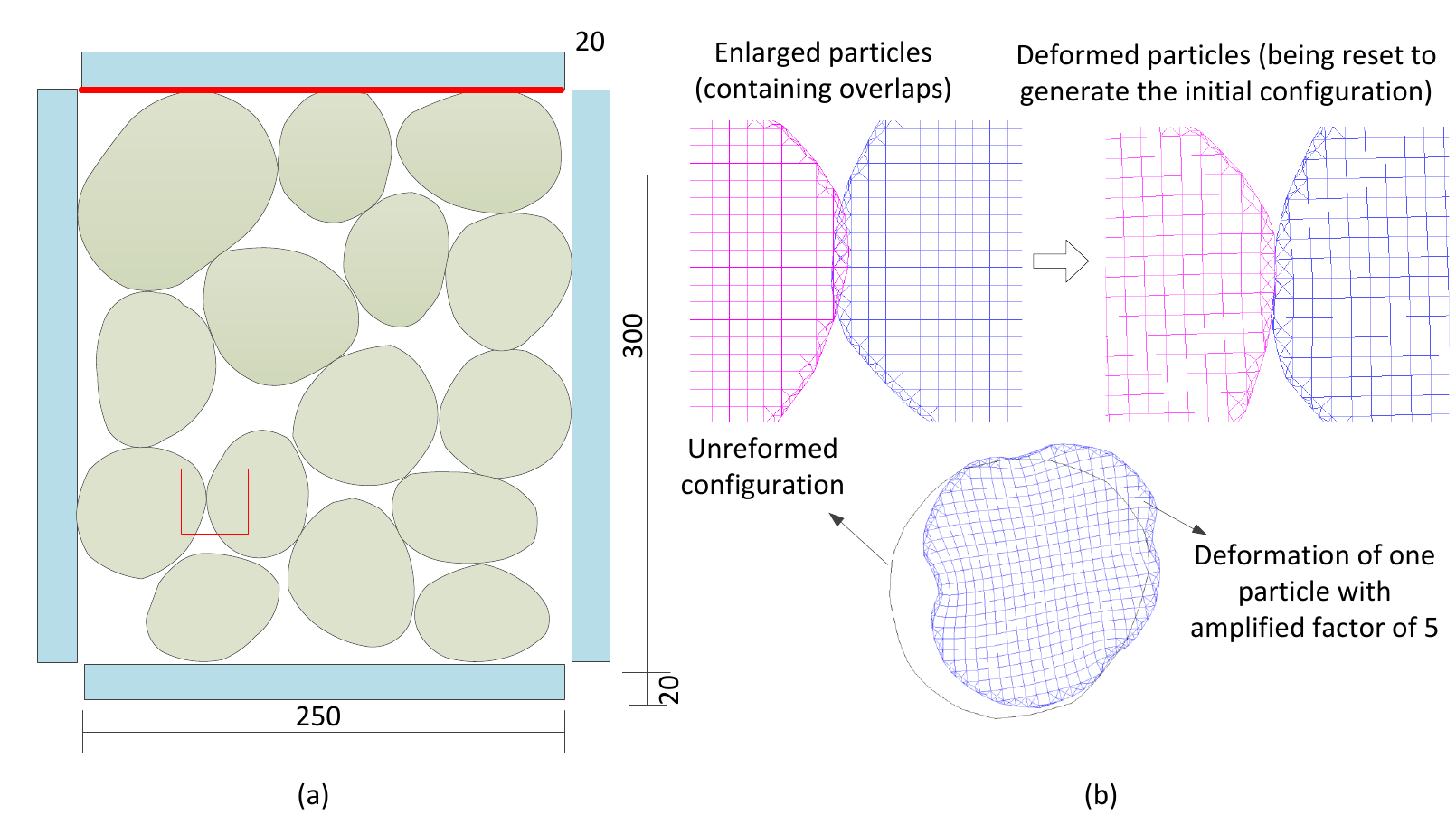}
\caption{Schematic of the model containing 15 particles: (a) configuration published in \cite{andrade2012granular} (the red line represents the cut surface to compute tractions) and (b) the process to generate the initial configuration.}
\label{case_fifteen_grains_model}
\end{figure}

We last study a grain assembly consisting of multiple non-sphere particles proposed in \citet{andrade2012granular}. As shown in Fig. \ref{case_fifteen_grains_model}a, fifteen deformable bodies are confined by four rigid plates with a uniform thickness of 20 mm. The Young's modulus is $E=100$ GPa, the Poisson's ratio is $\nu=0.2$, and the friction angle is 15$^{\circ}$. Note that we cannot exactly replicate the geometry shown in \citet{andrade2012granular} since we only redraw the particles by handle based on their published figure. If there exists an isolated particle without any overlap with neighboring particles in the redrawn figure, we will confront convergence issues due to the implicit algorithm. To avoid this problem, we generate the initial configuration as follows. First, we enlarge the grains in the redrawn figure by setting the level set of the boundaries as $\Phi = 0.5h$ (rather zero), where $h$ is the mesh size, to ensure that there are overlaps between bodies for convergence. We then fix the plates and conduct a simulation to achieve the equilibrium state. The deformed configuration is set as the initial configuration and stresses and displacements of the deformed particles are reset to zero. Fig. \ref{case_fifteen_grains_model}b shows this procedure. We also show the deformation of one particle in \ref{case_fifteen_grains_model}, where we emphasize that the deformation of particles may cause a convex particle becoming non-convex locally. This is particularly likely to happen on contacts of particles of different sizes or when the contact surface area is small for a given force. When the particle is nonconvex, the traditional DEM computing the contact force via the overlap may confront challenges (also for Fig. \ref{case_four_grains_results}). Therefore, we require a mean to consider the deformation.


\begin{figure}[h]
\centering
\includegraphics[width=9cm]{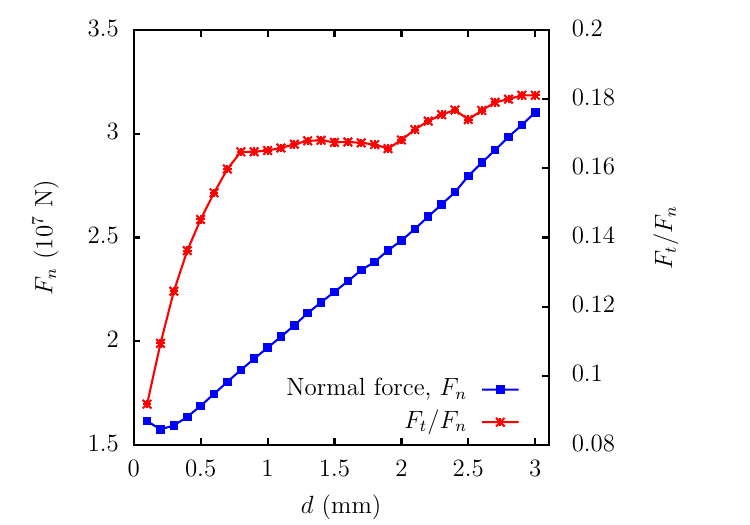}
\caption{Loading curves for the isotropic compressing phase}
\label{case_fifteen_particles_curve}
\end{figure}

\begin{figure}[h]
\centering
\includegraphics[width=15cm]{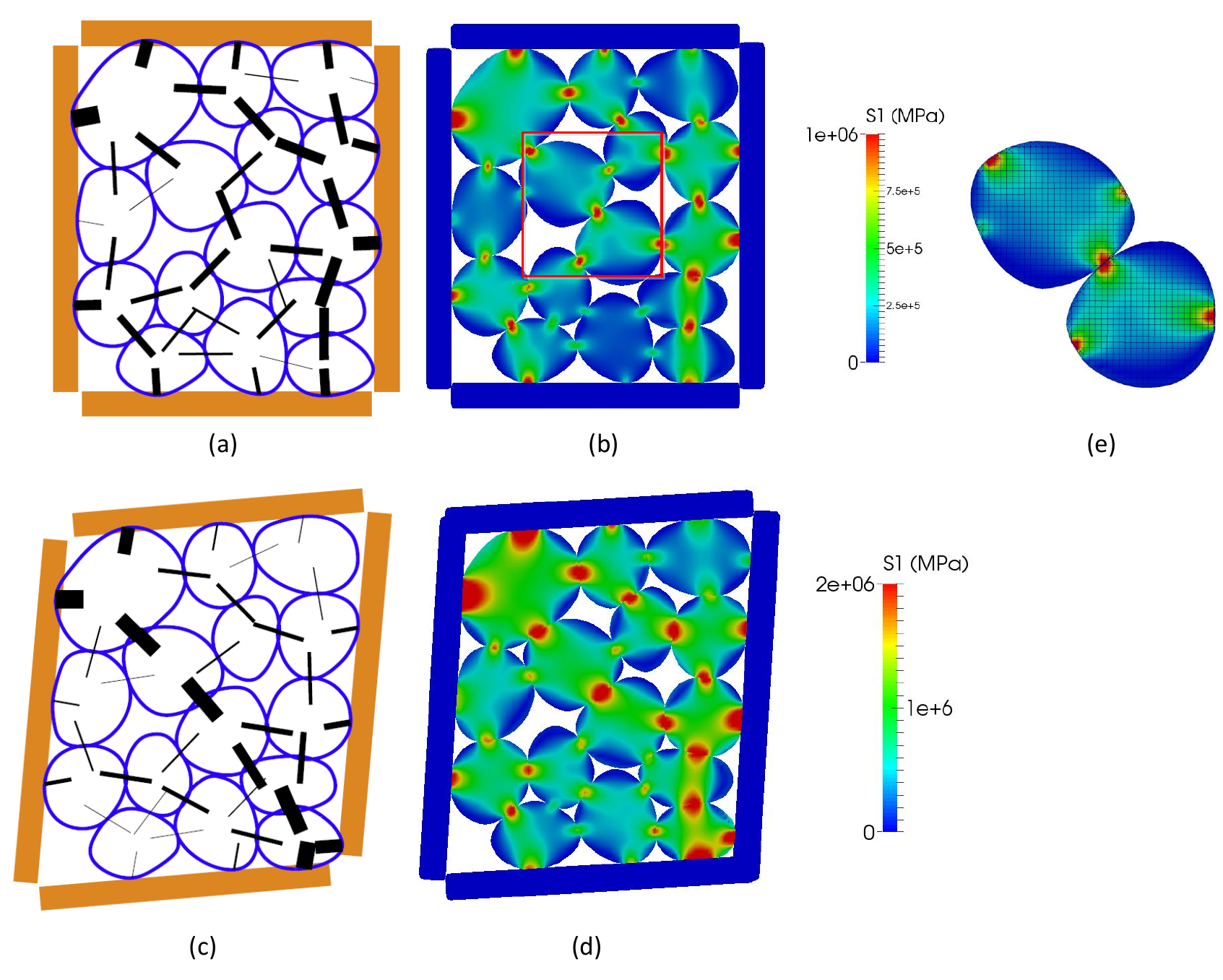}
\caption{Comparison of the results obtained from the proposed method and the DEM \citep{andrade2012granular}: (a) force chain obtained by the DEM for the compression, (b) distribution of the maximum stress obtained by the proposed method for the compression, (c) force chain obtained by the DEM for the pure shearing, (d) distribution of the maximum stress obtained by the proposed method for the pure shearing, and (e) details of the contact between two particles.}
\label{case_fifteen_particles_results}
\end{figure}

In this example, we first isotropically compress the sample by moving the plates toward the center with an incremental displacement $\Delta d=0.1$mm. Fig. \ref{case_fifteen_particles_curve} shows the evolution of the normal force and the ratio of tangential force and normal force within 30 loading steps. The normal and tangential forces are computed by a summation of contact forces for the contacts involving the top plate, which is approximately equivalent to the forces acting on the cutting plane as denoted by the red line in Fig. \ref{case_fifteen_grains_model}. We observe that the normal force linearly increase with the loading, but the ratio of the tangential force with the normal force reaches a steady state after several steps due to the deformation of the bodies. Fig. \ref{case_fifteen_particles_results}(a-b) compare the force chain obtained by the DEM \citep{andrade2012granular}. We can see that the general network of force chains are almost identical, but with some discrepancies. Fig. \ref{case_fifteen_particles_results}(e) shows details of two contacting bodies.

In the shear phase, we apply a shear loading on the sample by rotating the plates around their vertices, following the treatment in  \citet{andrade2012granular}. Note that this loading is not a simple shear load due to  the volume changes imposed by the boundary condition.
Nevertheless, this does not affect our purpose, which is to compare the results between the DEM and the proposed model.
Fig. \ref{case_fifteen_particles_results}(c-d) shows this comparison. Since the particles are extensively compressed, the stresses at the regions far away the contacting points are also very large, which cannot be reflected by the DEM.

\begin{figure}[h]
\centering
\includegraphics[width=16cm]{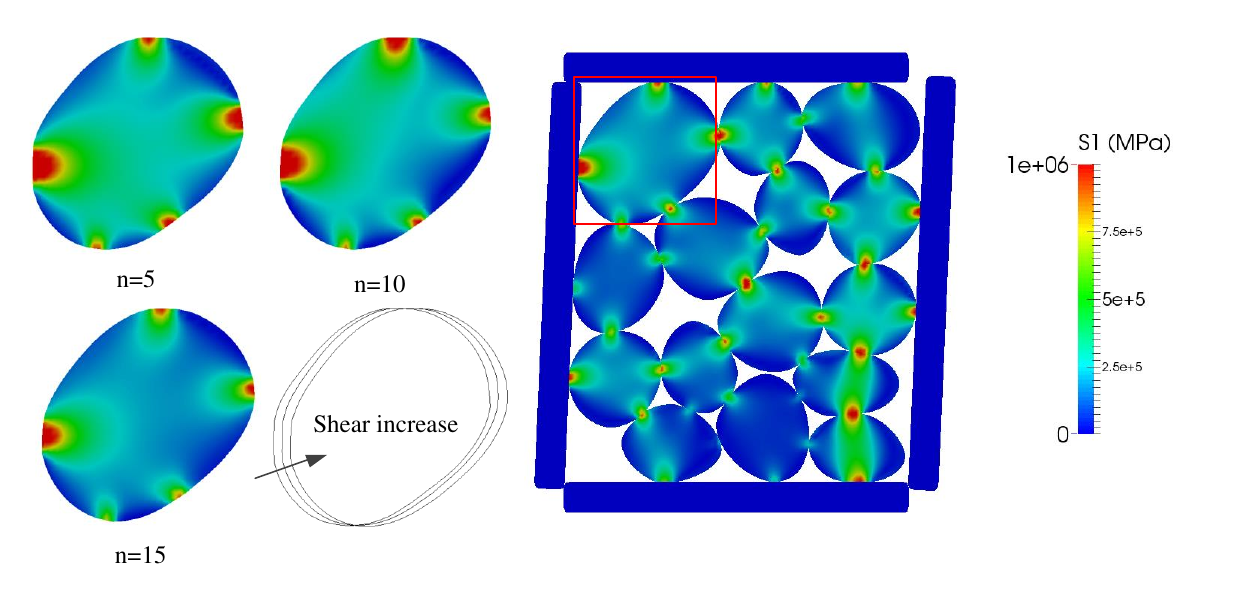}
\caption{Distribution of the maximum principle stress for the simple shear loading. The left subfigures are these distributions for a specific particle and its boundaries at different steps.}
\label{case_fifteen_particles_evolutin}
\end{figure}

\begin{figure}[h]
\begin{minipage}[c]{0.5\textwidth}
\centering
\includegraphics[width=1.1\textwidth]{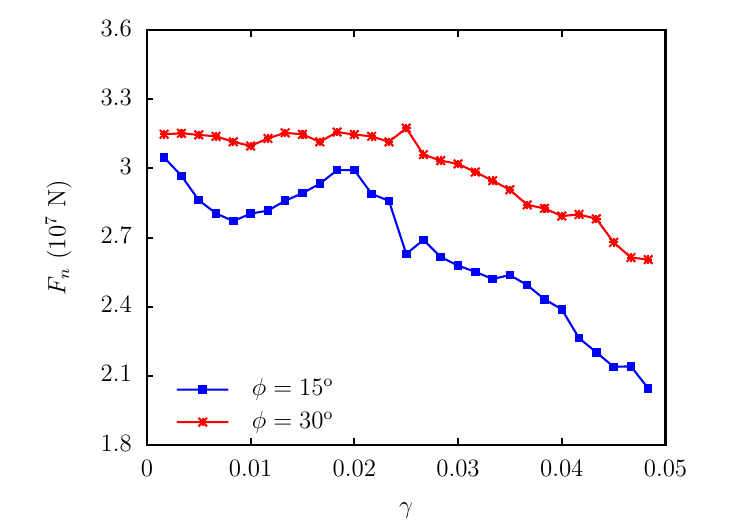}
\end{minipage}
\begin{minipage}[c]{0.5\textwidth}
\centering
\includegraphics[width=1.1\textwidth]{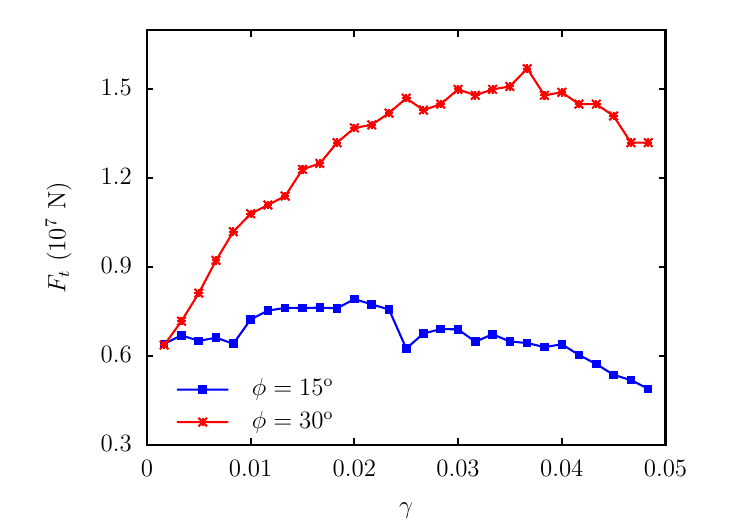}
\end{minipage}
\caption{Evolutions of normal and tangential forces for the simple shear test with different friction angles.}
\label{case_fifteen_particles_simpleshear}
\end{figure}

To explore the versatility of the proposed method, we conduct a simple shear of the sample after the compression phase by fixing the top and bottom plates and rotating the lateral plates with an incremental rotation angle of $0.3/180$ rad. We simulate 30 steps for the simple shear test. For the isotropic compression and the pure shearing, to some extent, the structures of the particles are stable resulting in continuous loading curves as shown in Fig. \ref{case_fifteen_particles_curve}. However, for the simple shear, we frequently observe  rearrangements of particles and reconstructions of the force chains. Fig. \ref{case_fifteen_particles_evolutin} shows the distribution of the maximum principle stress at the 25-th step. The left insets represent the evolutions of the stresses and boundaries of a specific particle. We can see that the stresses vary with the shear. The evolutions of normal and tangential forces are plotted in Fig. \ref{case_fifteen_particles_simpleshear}. Since the number of particles are only fifteen, the normal contact force cannot maintain constant for the conserve-volume simple shear test. Also, as discussed before, the curves are more fluctuated than the case of isotropic compression. We conducted another simulation with a different friction angle $30^{\circ}$ for comparison. As shown in Fig. \ref{case_fifteen_particles_simpleshear}, the shear force increases with the shear for large friction angle. 

Conclusively, we can capture the macroscopic responses of the assembly of the non-spheres and also obtain the field information for each particle, which provides a mean to record the local damage or plastic deformation of the particles. This would be our future work. 

\subsection{Conclusions}
In this work, we model the multi-body frictional contacts using level sets in voxel meshes. The boundaries of the bodies are implicitly represented by level sets and an unbiased contact reference is constructed to define the gap functions. The Coulomb friction law is analogized to an elastoplastic constitutive law to implement the regular return-mapping algorithm to unify the treatment for slip and stick states. We store the information of bodies by a set of material points and the level set variable is stored at the background mesh. After each step, both the material points and level sets are updated. The information of material points are reconstructed to the traditional Gauss's quadrature points of the elements discretized the updated boundaries. Examples are given to show the verifications and effectiveness of the proposed method to handle contacts involving multiple bodies and multiple loading steps.

In this work, we bypass the generations of conformal meshes and constructing cumbersome master-slave contact pairs. By using the voxel-based meshes, the number of DOFs dramatically reduces. Compared to the DEM, this method provides more information for the frictional contact problems. Since we know the stress field, in the future, we will consider the damage or crush for the particles. Thus, the macroscopic loading-curves would be more accurate. We aim to establish a bridge between real grain images and macroscopic constitutive laws.

\subsection{Acknowledgments}
This research is supported by
the Nuclear Energy University program from the Department
of Energy under grant contract DE-NE0008534,
the Earth Materials and Processes
program from the US Army Research Office under grant contract
W911NF-18-2-0306, the Dynamic Materials
and Interactions Program from the Air Force Office of Scientific
Research under grant contract FA9550-17-1-0169,
 as well as
the Mechanics of Materials and Structures program
at National Science Foundation under grant contract CMMI-1462760
and the NSF CAREER grant CMMI-1846875.
These supports are gratefully acknowledged.
The views and conclusions contained in this document are those
of the authors, and should not be interpreted as representing
the official policies, either expressed or implied, of the sponsors,
including the Army Research Laboratory or the U.S. Government.
The U.S. Government is authorized to reproduce and distribute
reprints for government purposes notwithstanding
any copyright notation herein.

\newpage

\bibliographystyle{plainnat}
\bibliography{main}

\end{document}